\documentclass[%
 reprint,
preprintnumbers,
 amsmath,amssymb,
 aps,
 pre,
]{revtex4-2}

\usepackage{graphicx}
\usepackage{dcolumn}
\usepackage{bm}

\usepackage{amssymb}
\usepackage{algorithm}
\usepackage{algpseudocode}
\usepackage{subcaption}
\usepackage{makecell}

\begin{document}

\preprint{APS/123-QED}

\title{A second-order volumetric boundary treatment for the lattice Boltzmann method}

\author{Kaj Hoefnagel}
\email{K.N.Hoefnagel@tudelft.nl}
\author{Damiano Casalino}
\author{Steven Hulshoff}
\author{Frits de Prenter}
\email{F.dePrenter@tudelft.nl}
\affiliation{%
Department of Flow Physics and Technology, Delft University of Technology, Kluyverweg 1, 2629HS Delft, Netherlands
}%

\date{\today}

\begin{abstract}
A new volumetric-type boundary treatment is introduced for the lattice Boltzmann method. Populations are projected onto a discontinuous piecewise linear basis and streamed using an exact geometrical mapping. The method is implemented in 2D for convex, stationary geometries. Stability is verified through a novel analysis tool based on the eigenvalues of the streaming operation. A convergence rate of $\geq2$ is achieved, indicating improved accuracy over precursor methods. For flow around a 2D airfoil, the method yields better lift and drag predictions.
\end{abstract}

\maketitle

\section{\label{sec:introduction}Introduction}
Historically, engineering problems governed by the Navier-Stokes equations have been simulated by discretizing these in space and time directly. The lattice Boltzmann method (LBM) has recently gained popularity as an alternative. It can be shown with a perturbation analysis that the weakly compressible Navier-Stokes equations emerge from the lattice Boltzmann equation if the continuum assumption holds \citep[Sec.~4.5.5]{Kruger2017}. The LBM offers several advantages: a local and explicit algorithm that gives superior parallel scaling \cite{Schornbaum2016}, low numerical dissipation \cite{Suss2023}, and relatively straightforward treatment of complex geometries \cite{Barad2017}. On the other hand, the LBM also has disadvantages due to its reliance on an equidistant Cartesian lattice. These include its inability to use anisotropic refinement to decrease the cost of resolving boundary layers and its need for complex algorithms to treat cells cut arbitrarily by boundaries. Proper enforcement of boundary conditions in cut cells remains problematic.

In general, the following five characteristics are desirable for an LBM boundary treatment: \textbf{1)} At least second-order convergence of the error, ensuring compatibility with the typically second order accuracy of the bulk \citep[p.~160]{Kruger2017}\cite{Sterling1996}. \textbf{2)} Exact conservation of mass and momentum, essential for ``fully self-consistent and stable microdynamics''\cite{Chen1998a}. \textbf{3)} Stability (implies that modes are never amplified by the boundary treatment). \textbf{4)} Smooth surface forces (which is an issue for Cartesian mesh methods in general \cite{dePrenter2018}). \textbf{5)} Computational cost, which should not exceed that of the bulk.

Three main categories of boundary treatments in the LBM exist:

\textbf{I) Broken-link-type schemes}, where missing populations on lattice links intersected by the boundary are found through inter-/extrapolation \cite{Bouzidi2001, Filippova1998, Mei1999, ZhaoLi2002, Chun2007}. These can be second-order schemes \cite{Mei1999}, but they are not necessarily conservative \citep[p.~45]{Marson2022}. Their stability can also be an issue, especially with high-order schemes at high Reynolds numbers \cite{Mei1999, DeRosis2014}. These schemes typically do not lead to smooth surface forces due to discrete sampling of the boundary \cite{Chen2012}, but their computational cost is typically low \cite{Nash2014}.

\textbf{II) Schemes based on the immersed boundary method (IBM)}, which impose the desired boundary condition through a force applied to nearby nodes \cite{Feng2004}. These schemes require interpolation, which can blur the boundary, leading to an incorrect apparent boundary location \cite{Breugem2012}. Furthermore, these schemes are generally only first-order accurate, and are usually not exactly conservative \citep[p.~473]{Kruger2017}. However, they are normally stable \cite{Peng2019}. Their surface forces are also not smooth \cite{Wu2009}, but their computational cost is typically low \cite{Feng2004}.

\textbf{III) Volumetric-type schemes} are based on a finite-volume-like approach. These geometrically compute the volume fraction of populations in a cell that will collide with a boundary segment. Incoming populations are collected for each boundary segment in the gather operation. These are then reversed and distributed back to the cells in the scatter operation. In the original scheme \cite{Chen1998b}, populations were assumed to be equally distributed throughout the cell, which led to first-order accuracy \cite{Li2012}. To address this, Li et al. later introduced a scattering correction \cite{Li2004, Li2009}. Li observes an overall order of convergence close to two, with the exception of boundary cells with small fluid fractions, where the order is close to one \cite{Li2012}. Next, since the full geometry is sampled, mass and momentum are exactly conserved. The method has good numerical stability \citep[p.~10]{Li2012}. Surface forces are more naturally computed due to the boundary-based approach \cite{Chen1998b}. However, oscillations still occur even with corrections \cite{Casalino2022}. Finally, the computational cost of this scheme has not been substantially addressed in the literature.

In this work, a new boundary treatment is introduced that is designed to satisfy all the aforementioned desired properties. We use an approach similar to volumetric-type schemes, but aim to achieve second-order accuracy by replacing the gather/scatter operation with an exact geometric mapping of populations reflecting off the boundary, and by representing populations with discontinuous piecewise polynomials within each cell. We refer to this treatment as the Continuum Field Formulation (CFF), emphasizing its use of spatial fields to represent the advection dynamics near the boundary.

We describe the performance of the CFF in terms of accuracy, stability, conservation, smoothness of surface forces, and computational cost.

First, an implementation of CFF in 2D is described in Sec.~\ref{sec:methodology}. Its stability is then analyzed using a novel approach based on the eigenvalues of the global streaming operation, described in Sec.~\ref{sec:Eigenvalues}. Next, the order of convergence is examined for a 1D and a 2D channel in Sec.~\ref{sec:ConvergenceResults}. Furthermore, results for a 2D airfoil are used to investigate the smoothness of surface forces. Finally, a first estimation of computational cost is provided in Sec.~\ref{sec:compCost}.

\section{Methodology}\label{sec:methodology}

\subsection{Fundamental LBM equations}
Rather than velocity and pressure, the LBM models the evolution of the particle distribution function $f_i(\mathbf{x}, t)$, which represents the density of particles moving with discrete velocity $\mathbf{c}_i$ at position $\mathbf{x}$ at time $t$. This function is hereafter referred to as the populations. In this work, the D2Q9 velocity model is used with $i\in \left\{0, 1, ..., 8\right\}$ and

\begin{equation}
\mathbf{c}_i =
\begin{cases} 
(0, 0) & i = 0, \\
(\pm 1, 0), (0, \pm 1) & i = 1, 2, 3, 4, \\
(\pm 1, \pm 1) & i = 5, 6, 7, 8,
\end{cases}
\label{eq:ci}
\end{equation}
where we nondimensionalize using lattice units, setting $\Delta x = 1$; $\Delta t = 1$.

The evolution equation of the LBM is given by:
\begin{equation}
    f_i(\mathbf{x} + \mathbf{c}_i\Delta t, t + \Delta t) = f_i(\mathbf{x}, t) + \Omega_i(\mathbf{x}, t),
\label{eq:evolution}
\end{equation}
where $\Omega_i$ is the collision operator. In this work, the BGK approximation is used for $\Omega_i$ \cite{Bhatnagar1954}:
\begin{equation}
    \Omega_i = -\frac{1}{\tau} \left( f_i - f_i^{\text{eq}} \right),
\end{equation}
where $\tau$ is the nondimensional relaxation time and $f_i^{\text{eq}}$ is the discrete equilibrium distribution \cite{Kruger2017}:

\begin{equation}
    f_i^{\text{eq}} = w_i \rho \left[ 1 + \frac{\mathbf{c}_i \cdot \mathbf{u}}{c_s^2} + \frac{(\mathbf{c}_i \cdot \mathbf{u})^2}{2c_s^4} - \frac{\mathbf{u} \cdot \mathbf{u}}{2c_s^2} \right].
\end{equation}
Here, $c_s=1/\sqrt{3}$ is the nondimensional speed of sound and $w_i$ are the weights which are $w_0=4/9$, $w_{1-4}=1/9$ and $w_{5-9}=1/36$ for the D2Q9 model. Furthermore, $\rho$ and $\mathbf{u}$ are the nondimensional macroscopic fluid density and velocity, respectively. They can be found as moments of $f_i$:
\begin{equation}
    \rho = \sum_i f_i, \qquad \rho \mathbf{u} = \sum_i f_i \mathbf{c}_i.
\label{eq:moments}
\end{equation}

Note that for some cases, we impose a gravity force
$\mathbf{g}$, in which case Eq.~\ref{eq:moments} and Eq.~\ref{eq:evolution} are modified as per the forcing scheme of Guo et al.\cite{Guo2002}:

\begin{equation}
    \rho \mathbf{u} = \sum_i f_i \mathbf{c}_i + \frac{\mathbf{g}}{2},
\end{equation}

\begin{equation}
    f_i(\mathbf{x} + \mathbf{c}_i\Delta t, t + \Delta t) = f_i(\mathbf{x}, t) + \Omega_i(\mathbf{x}, t) + S_i\left(\mathbf{x}, t\right),
\end{equation}

\noindent
where $S_i$ is the source term:

\begin{equation}
    S_i = \left(1 - \frac{1}{2\tau}\right) w_i \left[\frac{\mathbf{c}_i - \mathbf{u}}{c_s^2} + \frac{\left(\mathbf{c}_i\cdot \mathbf{u}\right)\mathbf{c}_i}{c_s^4}\right]\cdot\mathbf{g}.
\end{equation}

Finally, the nondimensional kinematic viscosity $\nu$ can be related to the relaxation time $\tau$ as:
\begin{equation}
    \nu = c_s^2\left(\tau - \frac{1}{2}\right).
\end{equation}

The evolution equation (\ref{eq:evolution}) is usually solved in two steps: \textbf{1) collision}, where $\Omega_i(\mathbf{x}, t)$ is added to $f_i(\mathbf{x}, t)$ and \textbf{2) streaming}, where populations advect from $\mathbf{x}$ to $\mathbf{x}+\mathbf{c}_i\Delta t$. The boundary treatment affects only the streaming step.

\subsection{Volumetric boundary treatments}
The LBM uses a Cartesian equidistant lattice which is usually represented by nodes in space, as in finite difference methods. Volumetric boundary schemes slightly modify this by using a finite-volume-like approach in the vicinity of boundaries, but only for the streaming step \cite{Chen1998b}. This approach is also used here. 

\begin{figure}
    \centering
    \includegraphics[width=0.7\linewidth]{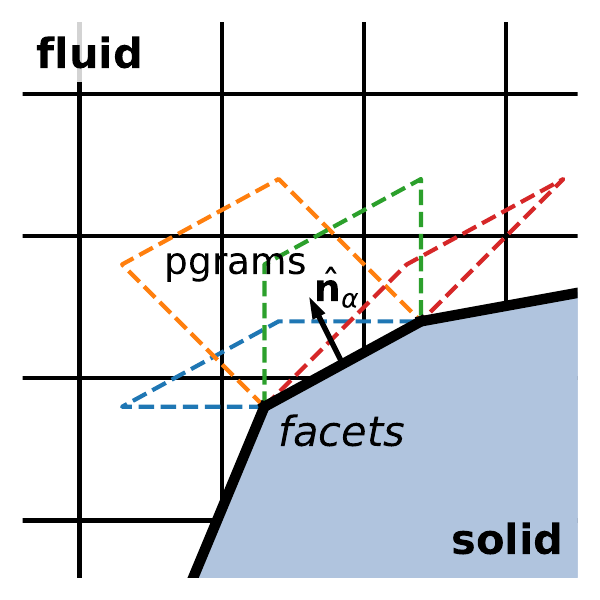}
    \caption{Facet and pgram definition, based on \citep[Fig.~3.2]{Li2012}.}
    \label{fig:pgrams}
\end{figure}

In volumetric boundary schemes, general solid objects in 2D are approximated by polygons, which are a collection of line segments. Following Chen et al. \cite{Chen1998b}, these line segments are referred to as facets. A set of parallelograms (pgrams) is obtained by extruding these facets by $-\mathbf{c_i}\Delta t$ for each velocity $\mathbf{c}_i$ that points into the facet (i.e. $\mathbf{c}_i\cdot \hat{\mathbf{n}}_\alpha < 0$) \cite{Chen1998b}, as shown in Fig.~\ref{fig:pgrams}. Each pgram then represents the spatial region that contains populations with velocity $\mathbf{c}_i$ that will reflect off the facet $\alpha$ in the next time step. Populations that are not overlapped by a pgram stream directly instead, without reflecting.

The original volumetric scheme collects all populations in a pgram (known as gathering), assuming they are uniformly distributed over each cell. It then reverses their velocity ($\mathbf{c}_i \to -\mathbf{c}_i$), and evenly spreads them back over the same pgram (known as scattering). This means that the populations are averaged over each pgram. In reality, populations starting close to the facet will end up far from the facet at the end of the time step and vice versa. Li et al. address this discrepancy through a scatter correction based on the local velocity gradient \cite{Li2004,Li2009}. However, their scheme still has an order of convergence of less than two \cite{Li2012}.

In this work, we represent populations $f_i$ using nonuniform, discontinuous piecewise polynomials. Furthermore, we remove the gather/scatter operations entirely and replace them with an exact geometric mapping. After the geometric mapping, $f_i$ is integrated over each cell, as in Chen et al. This is made numerically tractable using efficient quadrature routines developed for immersed methods \cite{dePrenter2023}. Algorithm~\ref{alg:CFF} summarizes the main steps. The geometric mapping as well as the construction of the discontinuous piecewise polynomials are detailed in the next two sections. In this study, we restrict ourselves to stationary, convex geometries, as handling concave or moving boundaries introduces additional complexities \cite{Chen1998b, Rohde2002}.

\begin{algorithm}[H]
\caption{Top-level LBM algorithm with proposed boundary treatment}
\label{alg:CFF}
\begin{algorithmic}[1]
\State $f_i(\mathbf{x}, 0)$ = \textsc{initialize}($\mathbf{u}(\mathbf{x}, 0), \rho(\mathbf{x}, 0)$)
\For{t=0 to endTime}
\State $p_i(\mathbf{x}, t, d)$ = \textsc{getCoeffs}($f_i(\mathbf{x}, t)$) \Comment{See Sec.~\ref{subsec:piecewisePolynomials}}
\State $f_i(\mathbf{x}+\mathbf{c}_i\Delta t, t)$ = \textsc{stream}($p_i(\mathbf{x}, t, d)$) \Comment{See Sec.~\ref{subsec:MappingAndIntegration}}
\State $f_i(\mathbf{x}, t+1)$ = \textsc{collide}($f_i(\mathbf{x}, t)$, $\tau$) \Comment{Default collision}
\EndFor
\end{algorithmic}
\end{algorithm}

\subsection{Geometric mapping and integration}\label{subsec:MappingAndIntegration}

\subsubsection*{Geometric mapping and integration in 1D}
To explain the geometric mapping, we first consider the dynamics of a population that reflects off a no-slip wall. For the no-slip condition, momentum is exactly reversed at the facet. Defining $i^*$ as the population opposite to $i$ (i.e. $\mathbf{c}_{i^*} \equiv -\mathbf{c}_i$), we can correspondingly write: $f_{i}^{out}=f_{i^*}^{in}$ \cite{Chen1998b}. Geometric mapping then consists of three steps for populations $f_{i^*}$ inside the pgram: travel distance $d$ to the facet, reverse direction ($f_{i^*} \to f_i$), and travel distance $1-d$ away from the facet. With the geometric mapping, we want to capture these dynamics exactly at each point in the pgrams.

\begin{figure*}[t]
    \begin{subfigure}[t]{0.32\textwidth}
        \centering
        \includegraphics[width=\linewidth]{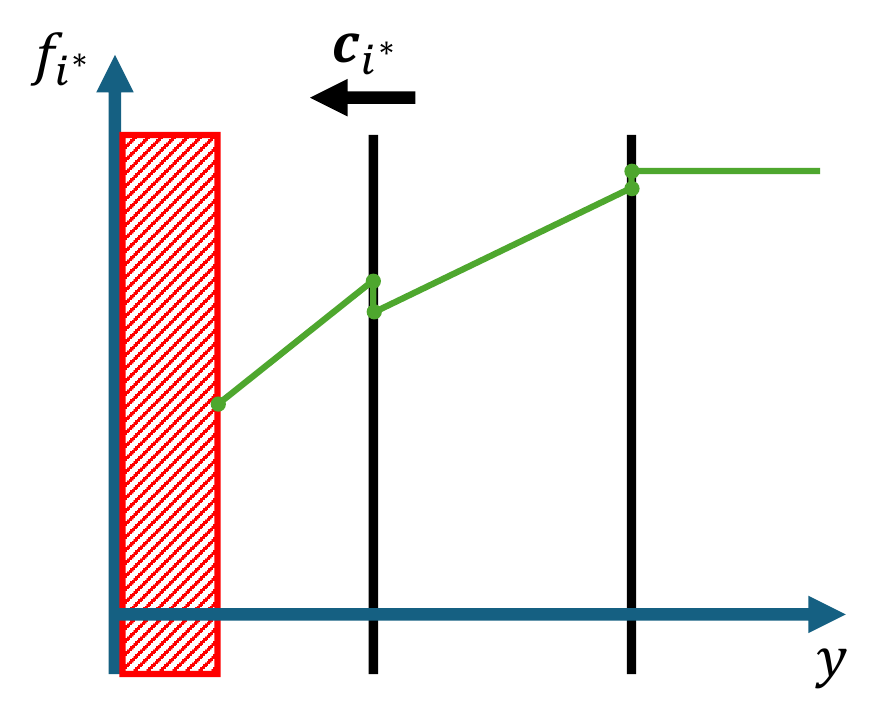}
        \caption{Pre-stream left-moving populations}
        \label{fig:1DMappingF3_prestream}
    \end{subfigure}
    \begin{subfigure}[t]{0.32\textwidth}
        \centering
        \includegraphics[width=\linewidth]{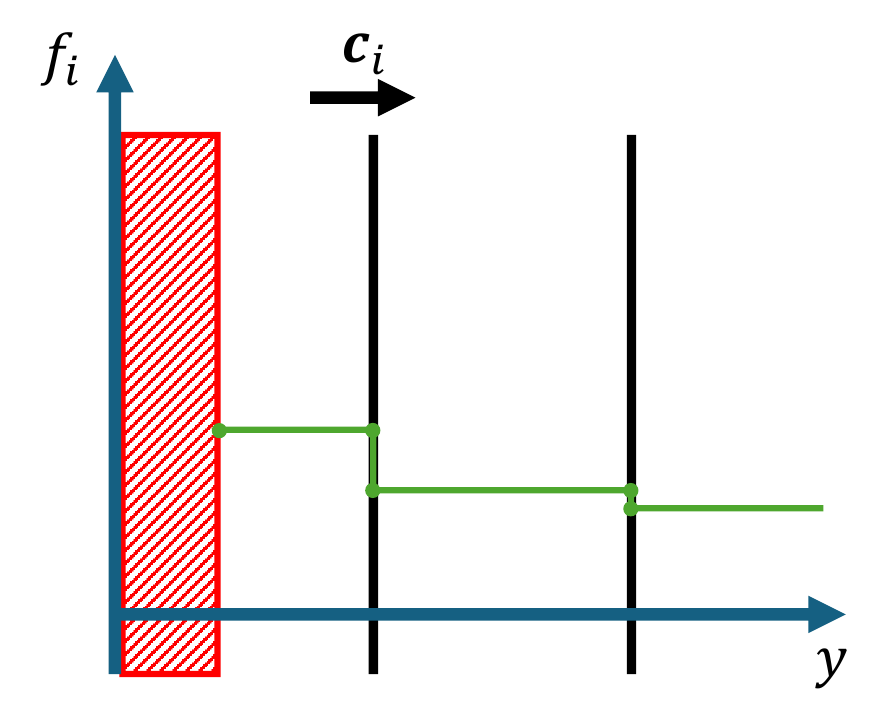}
        \caption{Pre-stream right-moving populations}
        \label{fig:1DMappingF1_prestream}
    \end{subfigure}
    \begin{subfigure}[t]{0.32\textwidth}
        \centering
        \includegraphics[width=\linewidth]{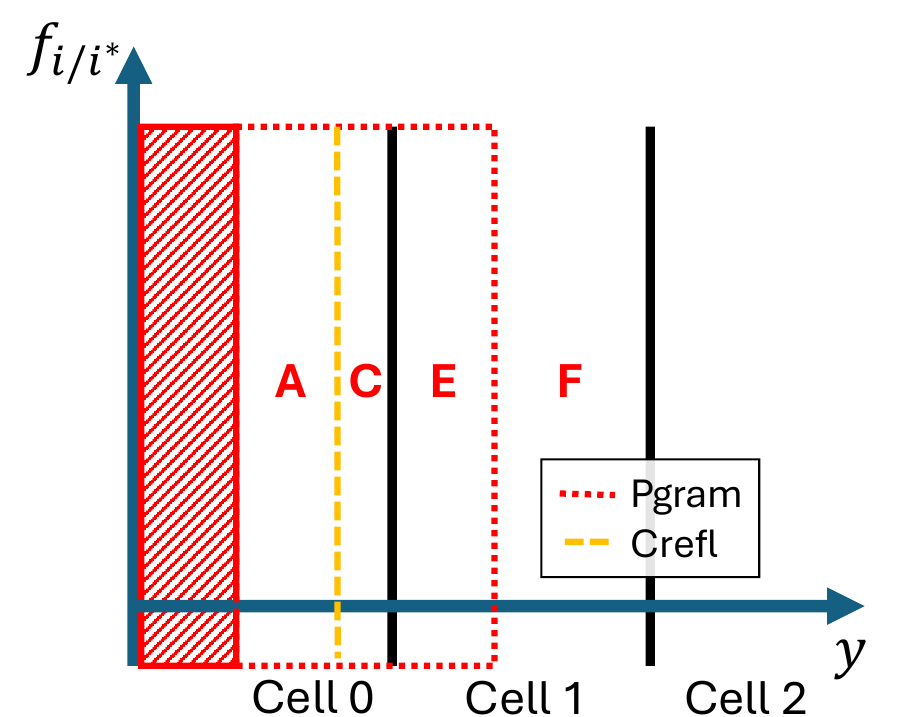}
        \caption{Geometric subzones over which populations are integrated before streaming}
        \label{fig:1DMapping_zones}
    \end{subfigure}
    \begin{subfigure}[t]{0.32\textwidth}
        \centering
        \includegraphics[width=\linewidth]{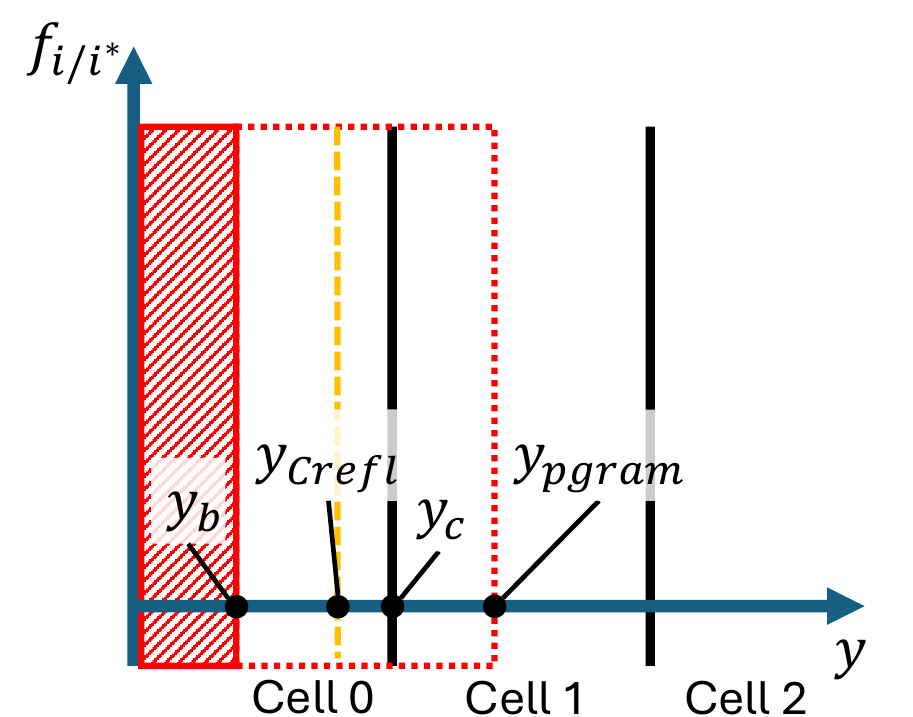}
        \caption{Coordinates of geometric subzones in Fig.~\ref{fig:1DMapping_zones}}
        \label{fig:1DMapping_coords}
    \end{subfigure}
    \begin{subfigure}[t]{0.32\textwidth}
        \centering
        \includegraphics[width=\linewidth]{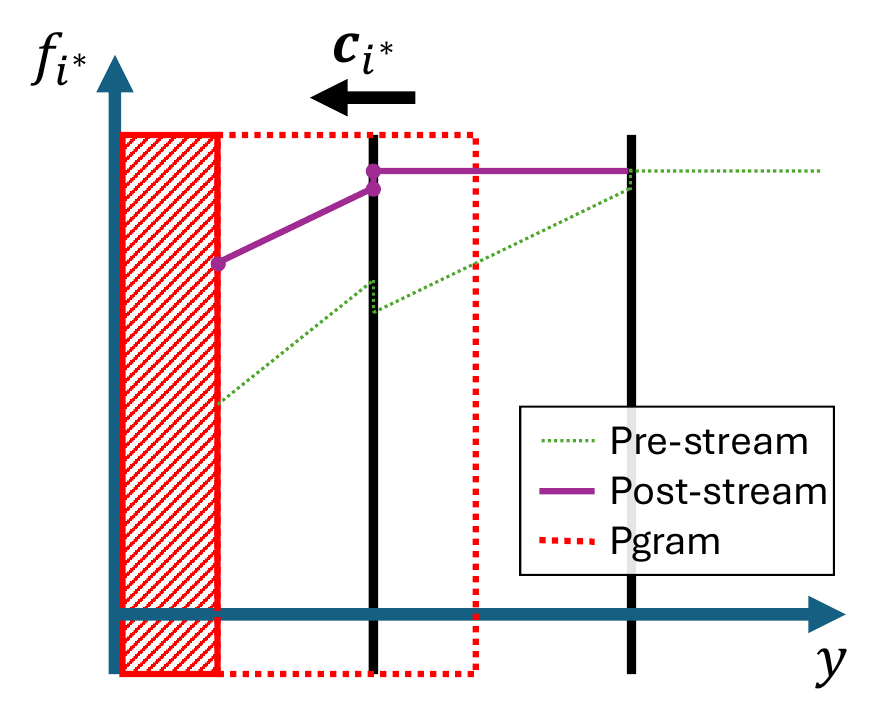}
        \caption{Post-stream left-moving populations}
        \label{fig:1DMappingF3_poststream}
    \end{subfigure}
    \begin{subfigure}[t]{0.32\textwidth}
        \centering
        \includegraphics[width=\linewidth]{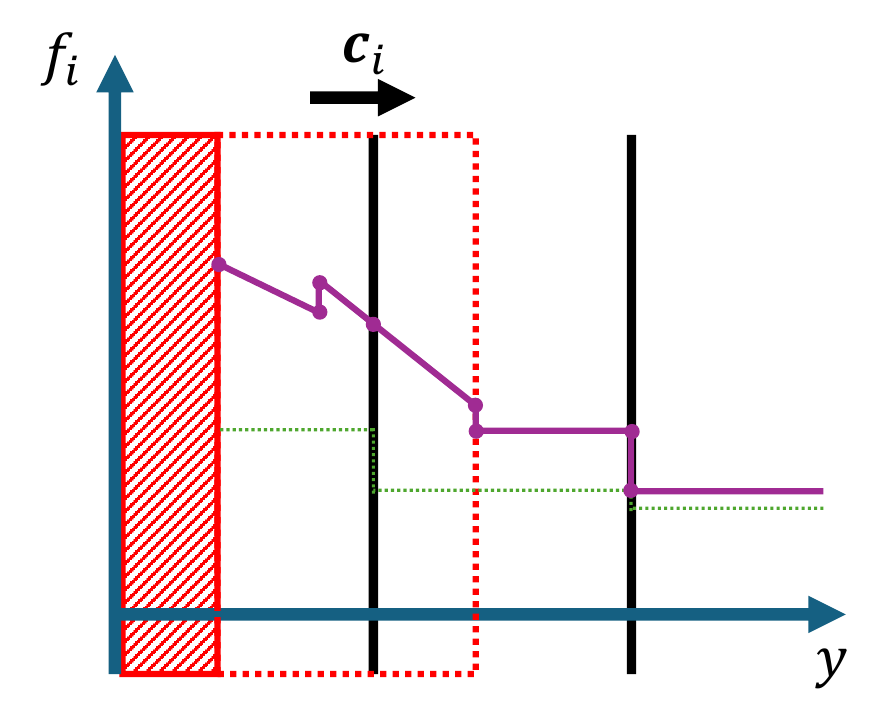}
        \caption{Post-stream right-moving populations}
        \label{fig:1DMappingF1_poststream}
    \end{subfigure}
    \caption{Illustration of the geometric mapping in 1D. Three cells are shown with the leftmost one cut by a boundary.}
    \label{fig:mapping1D}
\end{figure*}

The procedure will first be explained in 1D using Fig.~\ref{fig:mapping1D}. First, discontinuous piecewise polynomials are used to get a representation of the populations at each point in space. The left- and right-moving populations before streaming are shown in Figs.~\ref{fig:1DMappingF3_prestream} and \ref{fig:1DMappingF1_prestream}. Then we stream and apply the aforementioned reflection dynamics to these discontinuous piecewise polynomials. This results in the post-stream populations shown in Figs.~\ref{fig:1DMappingF3_poststream} and \ref{fig:1DMappingF1_poststream}. 

After the geometric mapping, the post-stream populations in Figs.~\ref{fig:1DMappingF3_poststream} and \ref{fig:1DMappingF1_poststream} are integrated over the cells to obtain population values in lattice points. In our implementation, these integrals are realized by first integrating the pre-stream populations over subzones and then applying geometric mapping to the integrated results. The corresponding subzones are illustrated in Fig.~\ref{fig:1DMapping_zones}, labeled A, C, E, and F. The bounds of these zones are defined such that each zone lies in only one cell and also maps to only one cell. Specifically, zone A is defined such that left-moving populations starting there reflect and exclusively flow to zone E. Similarly, left-moving populations in C reflect and exclusively flow to C and E to A. Populations outside the pgram stream normally. For example, populations in zone F will stream directly (without reflecting) to zone A or C. 

The coordinates of the edges of the subzones are indicated in Fig.~\ref{fig:1DMapping_coords}, they need to be computed to evaluate the integrals. In 1D, they are: 
\begin{itemize}
    \item \textbf{Facet location} $\mathbf{y}_b$: location of the boundary.
    \item \textbf{End of pgram} $\mathbf{y}_{\text{pgram}}$: located one lattice distance from the facet in direction $\mathbf{c}_i$; $\mathbf{y}_{\text{pgram}}=\mathbf{y}_b + \mathbf{c}_i$.
    \item \textbf{Cell borders} $\mathbf{y}_c$: located at each integer $y$-location; $\mathbf{y}_c=\pm k$, $k\in \mathbb{Z}$.
    \item \textbf{Reflected cell border} $\mathbf{y}_{\text{Crefl}}$: This border separates populations that will reflect off the facet and end up in cell 1 from those that end up in cell 0. This means that a leftmoving population starting at this border would reflect off the facet and end up exactly at the cell border between cell 0 and 1 ($\mathbf{y}_c$). It thus represents the reflection of a cell border (Crefl). Its location $\mathbf{y}_{\text{Crefl}}$ is found using:
    \begin{equation}
        (\mathbf{y}_c - \mathbf{y}_b) + (\mathbf{y}_{\text{Crefl}} - \mathbf{y}_b) = \mathbf{c}_i.
        \label{eq:Crefl1D}
    \end{equation}
    Note that Eq.~\ref{eq:Crefl1D} implies that when the facet is in the right half of cell 0 instead of the left half, $\mathbf{y}_{\text{Crefl}}$ lies in cell 1. In this case the zones are defined slightly differently than those in Fig.~\ref{fig:1DMapping_zones}, but the concept is the same. Nonetheless, a distinction between these two cases has to be made explicitly in the algorithm. 
\end{itemize}

The final goal is to obtain integrals of the post-stream populations. Using the streaming relations between subzones, these are written as sums of pre-stream populations integrated over subzones. These sums are given in Tab.~\ref{tab:zoneIntegralRelations}. Finally, note that the integration of $f_i$ in space results in a number of particles, $N_i$. To convert $N_i$ back to a concentration $f_i$, it is divided by the cell volume, as further discussed in \cite{Chen1998b}.
\begin{table}
    \centering
    \caption{Expression of post-stream integrals over cells in terms of pre-stream integrals over subzones and cells. Cell integrals are only over the fluid portion.}
    \label{tab:zoneIntegralRelations}
    \begin{tabular}{|c|l|} \hline
    Post-stream integral & Pre-stream integral(s) \\ \hline
    $f_i$ over cell 0 & $f_{i^*}$ over C $+$ $f_{i^*}$ over E \\ \hline
    $f_i$ over cell 1 & $f_{i^*}$ over A $+$ $f_i$ over cell 0 \\ \hline
    $f_{i^*}$ over cell 0 & $f_{i^*}$ over F \\ \hline
    $f_{i^*}$ over cell 1 & $f_{i^*}$ over cell 2 \\ \hline
    \end{tabular}
\end{table}

Note that since $f_i(\mathbf{x})$ is represented with discontinuous piecewise polynomials, the integrals can be evaluated analytically. This is the method used in the present work. However, these integrals could also be calculated using advanced quadrature \cite{dePrenter2023} to reduce cost in more complex cases.

\subsubsection*{Geometric mapping and integration in 2D}
\begin{figure}
    \centering
    \includegraphics[width=\linewidth]{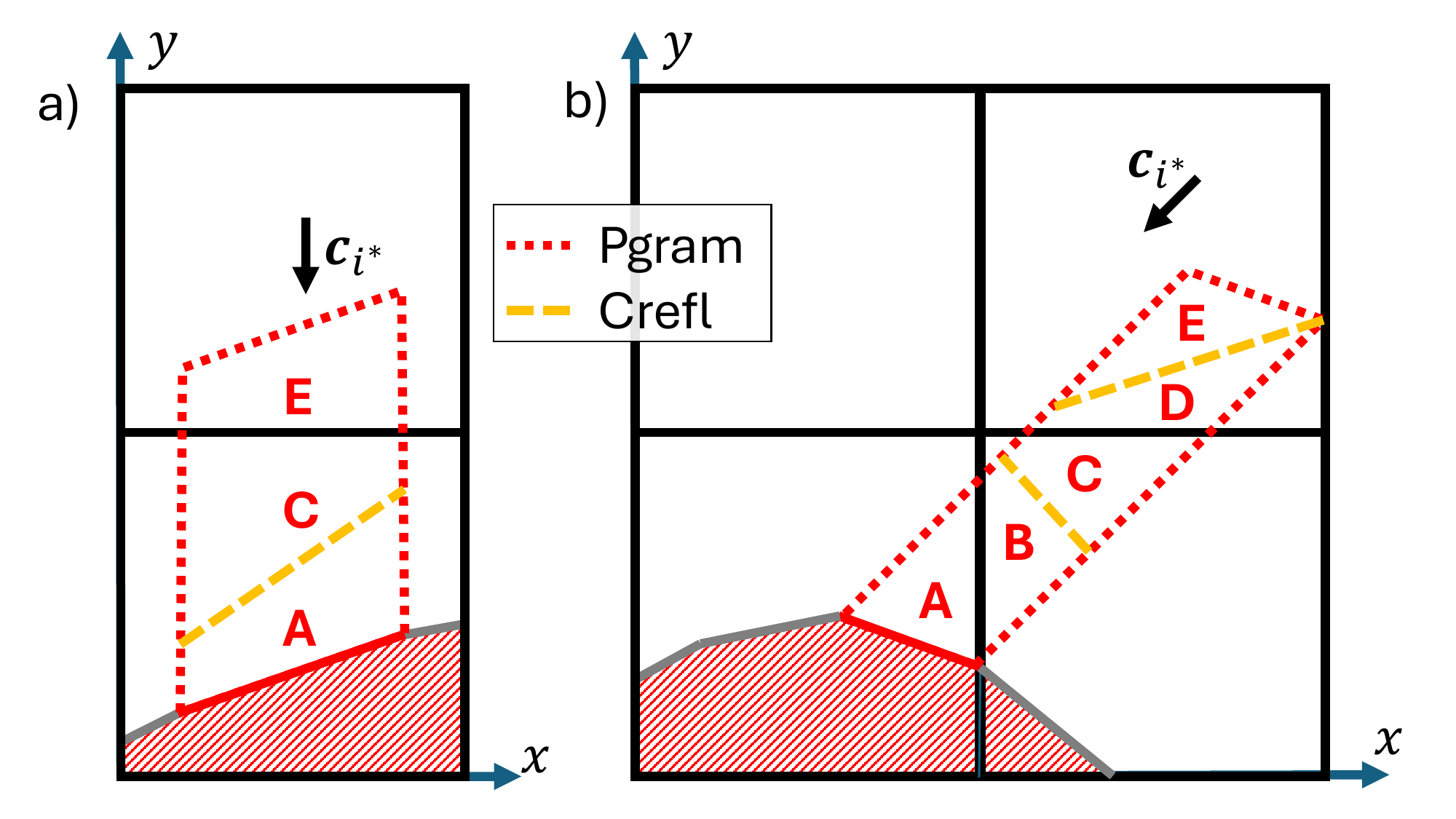}
    \caption{Pgram subzones for the geometric mapping of \textbf{a)} a 2D straight reflection ($\left|\mathbf{c}_i\right|=1$) and \textbf{b)} a 2D diagonal reflection ($\left|\mathbf{c}_i\right|=\sqrt{2}$).}
    \label{fig:mapping2D}
\end{figure}

For 2D geometric mapping, two cases are distinguished; a straight reflection ($\left|\mathbf{c}_i\right|=1$) and a diagonal reflection ($\left|\mathbf{c}_i\right|=\sqrt{2}$), they are shown in Fig.~\ref{fig:mapping2D}. The straight reflection can be thought of as a 1D reflection in $y$, extruded in the $x$-direction. The same pgram subzones are used which reflect in the same way (down-moving populations in A reflect and flow to E, similarly, C to C and E to A).  

For the diagonal reflection, the pgram intersects three cells, leading to two Crefl lines (one from the horizontal and one from the vertical grid line), this results in five subzones. Again, the populations $f_{i^*}$ in one subzone exclusively end up as populations $f_i$ in another subzone. Thus, down-left-moving populations in A reflect and flow to E. Similarly, B to D, C to C, D to B, and E to A.

\begin{figure}
    \centering
    \includegraphics[width=0.3\linewidth]{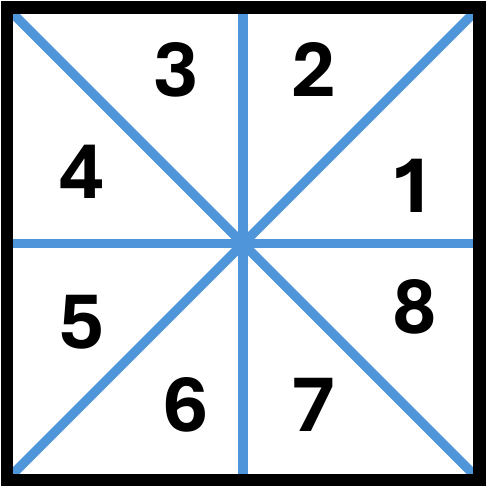}
    \caption{Cell regions which determine the exact configuration of the subzones of the diagonal reflection.}
    \label{fig:2DDiagZones}
\end{figure}

The particular cell in which each subzone lies depends on the location of the facet within its cell. For the vertical reflection, there are two options: the facet lies in the top half of the cell or the facet lies in the bottom half of the cell. For the diagonal reflection there are eight such options, corresponding to the geometric regions within the cell shown in Fig.~\ref{fig:2DDiagZones}. In the algorithm, facets are explicitly associated with one of these eight regions. Facets that span multiple regions are subdivided so that each only lies in one region.

Finally, integrals of post-stream populations need to be calculated. These are again expressed as sums of integrals of pre-stream populations over the subzones. In our implementation, we represent the boundary using straight facets, resulting in polygonal subzones. The integrals of the discontinuous piecewise polynomials over these polygons are evaluated analytically using the expression derived by C. Steger \cite{steger1996}. This requires the corner coordinates of each subzone in Fig.~\ref{fig:mapping2D}. The corners of the pgram and the intersection points between the pgram and cell borders are trivially found. To find the corners of the reflected cell border (Crefl), Eq.~\ref{eq:Crefl1D} is used again, where $\mathbf{y}_b$ are the endpoints of the facet and $\mathbf{y}_c$ the intersection points of the pgram and the relevant cell border (horizontal/vertical).

\subsection{Construction of the discontinuous piecewise linears}\label{subsec:piecewisePolynomials}
\begin{figure}
    \centering
    \includegraphics[width=0.8\linewidth]{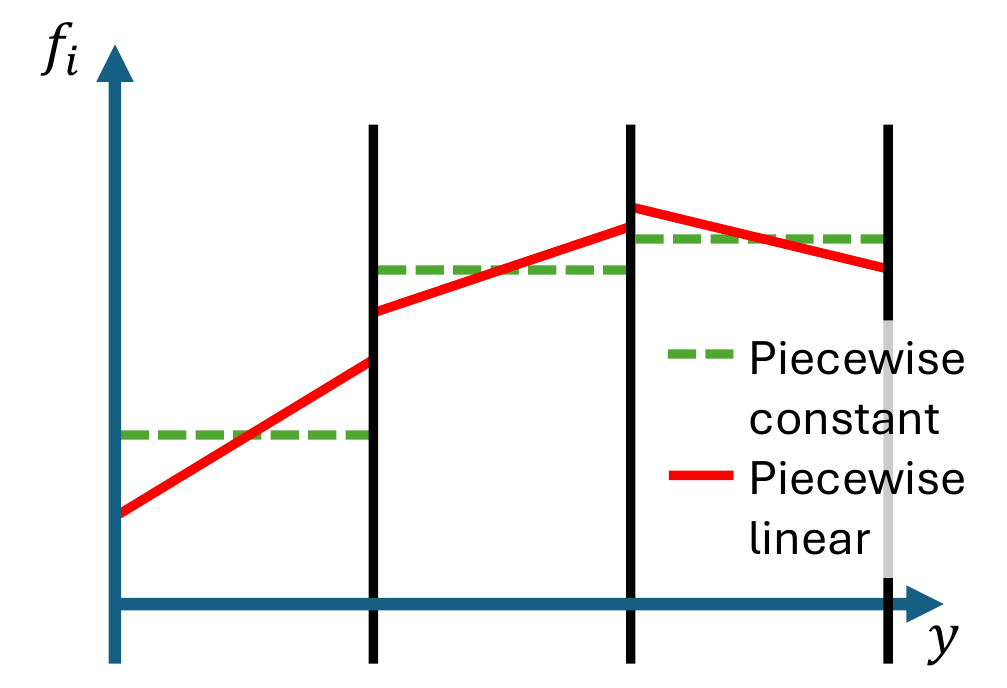}
    \caption{Piecewise constant and piecewise linear representations of $f_i(\mathbf{x}, t)$.}
    \label{fig:1DPiecewiseRepresentation}
\end{figure}

Following the streaming operation outlined in Fig.~\ref{fig:mapping1D}, $f_i$ must be integrated over arbitrary parts of cells. To represent $f_i$, we take a finite-volume-like approach, where instead of representing it in the lattice point, we distribute it over the cell. Note that this can be interpreted as a generalization of Chen et al. \cite{Chen1998b}, who distributed the populations over the cell as piecewise constant, as shown in Fig.~\ref{fig:1DPiecewiseRepresentation}, leading to a first-order accurate scheme \cite{Li2012}. In order to obtain a higher order of accuracy, we distribute $f_i$ using discontinuous piecewise (linear) polynomials per cell, also shown in Fig.~\ref{fig:1DPiecewiseRepresentation}. In the following, the procedure to construct these is first described for 1D and subsequently for 2D.

\subsubsection*{Construction of the discontinuous piecewise linears in 1D}\label{subsubsec:PP1D}
The input consists of the populations $f_i$ in the lattice points. These populations are physically located at the centroid of each cell \cite{Chen1998a}. For non-cut cells the centroid coincides with the lattice point (cell center), but for cut cells it differs. To obtain higher-degree polynomials, information from neighbours has to be used. Increasing the degree of the polynomials increases the number of neighbours needed and thus reduces the locality of the scheme (which we aim to preserve as much as possible). We observe that discontinuous piecewise linears (first-degree polynomials) already yield the desired second-order scheme, so these are used in this work. Nonetheless, generalization to higher-degree polynomials is possible. Construction of discontinuous piecewise linears of the form $f_{i^*}(y)=a+by$ requires two constraints. We use:
\begin{enumerate}
    \item Exact conservation of the populations in the cell.
    \item When the discontinuous piecewise linear is hypothetically extended to its neighbours, it should satisfy the conservation of populations in the neighbours as accurately as possible.
\end{enumerate}

\begin{figure}
    \centering
    \includegraphics[width=\linewidth]{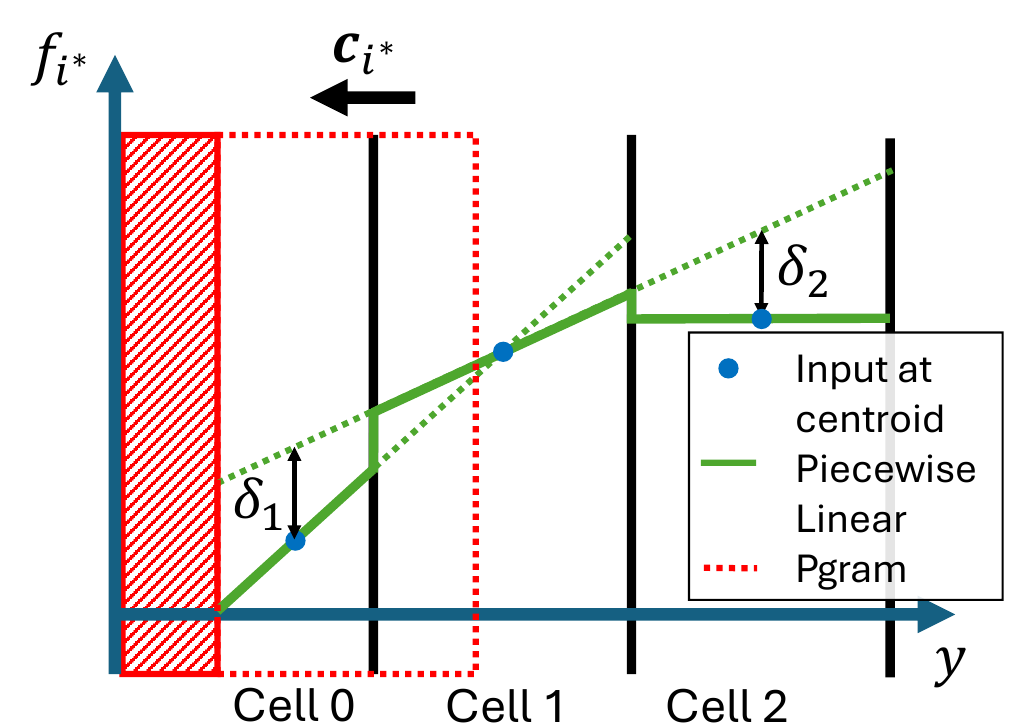}
    \caption{Construction of the discontinuous piecewise linears in 1D.}
    \label{fig:PP1D}
\end{figure}

Construction of the discontinuous piecewise linears based on these two criteria is illustrated in Fig.~\ref{fig:PP1D}. For linear functions, the first condition is satisfied simply when it attains the input value at the centroid of the cell. For the second condition, the extended piecewise linear should be as close as possible to the input value of the direct neighbours at their respective centroids. For cell 0, which only has one direct neighbour, this results in an exact fit through this neighbour's centroid. For cell 1, which has two direct neighbours, this results in an underdetermined system where $\delta_1$ and $\delta_2$ have to be minimized. We use a weighted L$_2$ minimization of $\delta_1$ and $\delta_2$. Various methods are possible:
\begin{itemize}
    \item Non-weighted: each neighbour cell has the same weight.
    \item Volume-weighted: each neighbour cell is weighted by their volume.
    \item Boundary-weighted: only pgram-overlapped cells are used; non-pgram-overlapped cells get a weight of zero.
\end{itemize}

In order to choose a weighting method, we first consider stability. Using the eigenvalue analysis which is further elaborated in Sec.~\ref{sec:Eigenvalues}, the boundary-weighted method was found to be unstable. We expect that this is caused by effectively approximating gradients with downwind information, considering that the populations are travelling towards the wall. On the other hand, the non-weighted and volume-weighted methods are both stable. Both have the same order of convergence, but the non-weighted method is a few percent more accurate. Nonetheless, the volume-weighted method is better supported by physical arguments. Namely, as a cell approaches zero volume, its influence on the solution should also vanish. Thus, we use the volume-weighted method in this work.

Note that the discontinuous piecewise linears need only to be constructed for populations flowing into the boundary within pgram-overlapped cells (only left-moving populations in cell 0 and 1 in Fig.~\ref{fig:PP1D}). Populations in other cells/directions exclusively stream to another cell, so their integral value does not depend on variations within the cell.

\subsubsection*{Construction of the discontinuous piecewise linears in 2D}\label{subsubsec:PP2D}
\begin{figure}
    \centering
    \includegraphics[width=\linewidth]{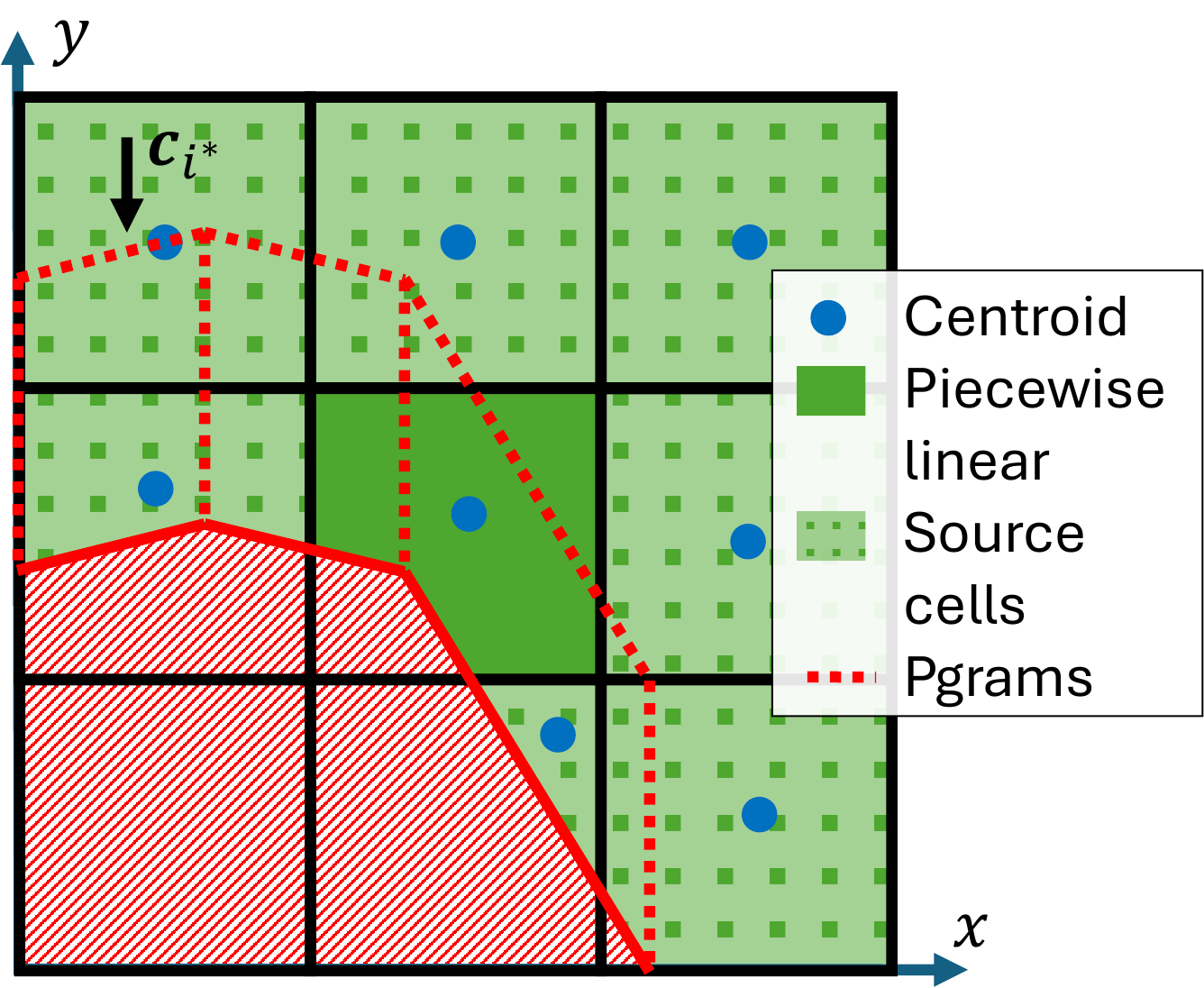}
    \caption{Construction of the discontinuous piecewise linear for the middle cell in 2D, using the surrounding source cells.}
    \label{fig:PP2D}
\end{figure}

In 2D, discontinuous piecewise linears are of the form $f_{i^*}=a + by + cx$. Three constraints are thus required to obtain the coefficients. Alternative linear bases also include a $xy$ component in this function. However, based on stability arguments which are further discussed in Sec.~\ref{sec:Eigenvalues}, we do not employ them. The same constraints are used as for 1D, but the L$_2$ minimization resulting from the second constraint is now solved for two unknowns. Up to eight direct neighbours can be used (four that directly border the cell and four that diagonally border the cell). An example is shown in Fig.~\ref{fig:PP2D}, where there are seven valid neighbours. For the minimization problem to be well posed, there should be at least two direct neighbours that are not colinear with the center cell (i.e. the three do not lie on one line). For convex boundaries this property is always satisfied. Finally, the same three weighting methods used in 1D can be extended to 2D. We again use volume-weighted for the results presented here.

\section{Eigenvalue analysis of the streaming operation}\label{sec:Eigenvalues}
In this section, a novel approach for the analysis of boundary treatments is introduced. It uses the eigenvalues associated with the global streaming operation to assess whether the boundary treatment is stable and to break down the error of the boundary treatment into a dissipative and a dispersive part. The concept is first explained in Sec.~\ref{subsec:conceptEigenValues}. Then, after a 2D testcase is introduced in Sec.~\ref{subsec:inclinedChannelSetup}, the results of a 2D analysis are presented in Sec.~\ref{subsec:eigenvalueResults}.

\subsection{Formulation}\label{subsec:conceptEigenValues}

\subsubsection*{The streaming matrix}
Construction of the discontinuous piecewise linears, as detailed in Sec.~\ref{subsec:piecewisePolynomials}, is a purely linear operation. Hence, it can be captured in a matrix-vector product:
\begin{equation}
    \mathbf{p} = \mathbf{R} \mathbf{f}^{\text{pre-stream}},
\label{eq:polynomialCoeffMatrix}
\end{equation}
where $\mathbf{f}^{\text{pre-stream}}$ is a vector with the pre-stream populations, $\mathbf{p}$ is a vector with the piecewise linear coefficients, and $\mathbf{R}$ is a matrix that encodes the calculation of the piecewise linear coefficients. 

The geometric mapping and subsequent cell integration, detailed in Sec.~\ref{subsec:MappingAndIntegration}, is also purely linear. It can be captured in a similar matrix-vector product:
\begin{equation}
    \mathbf{f}^{\text{post-stream}} = \mathbf{M} \mathbf{p},
\label{eq:mappingAndIntegrateMatrix}
\end{equation} 
where $\mathbf{f}^{\text{post-stream}}$ is a vector with the post-stream populations. Furthermore, $\mathbf{M}$ is a matrix that encodes: \textbf{1)} the streaming of the discontinuous piecewise linears (including the geometric mapping), \textbf{2)} the integration to number of particles in each cell ($N_i$), and \textbf{3)} the final conversion back to populations ($f_i$). 

Combining Eq.~\ref{eq:polynomialCoeffMatrix} and Eq.~\ref{eq:mappingAndIntegrateMatrix}:
\begin{equation}
    \mathbf{f}^{\text{post-stream}} = \mathbf{M}\mathbf{R}\mathbf{f}^{\text{pre-stream}} = \mathbf{S}\mathbf{f}^{\text{pre-stream}},
\end{equation}
where $\mathbf{S}=\mathbf{M}\mathbf{R}$ is the streaming matrix which encodes the full streaming step, including the enforcement of the boundary condition. 

\subsubsection*{Eigenvalues of the streaming matrix}
Analyzing the eigenvalues and eigenvectors of $\mathbf{S}$ yields insights into how the boundary treatment influences the amplification of $f_i$ during the streaming step. The corresponding eigenvalue problem is:
\begin{equation}
    \lambda_k \mathbf{v}_k = \mathbf{S}\mathbf{v}_k,
\end{equation}
where $\lambda_k$ and $\mathbf{v}_k$ are the $k$th eigenvalue and eigenvector, respectively. For relatively small cases, these can be found numerically. They will generally be complex numbers of the form $a_k + b_k\mathrm{j}$ with $\mathrm{j}=\sqrt{-1}$. Also, note that $\mathbf{S}$ is a block matrix, with separable blocks for each velocity pair $\mathbf{c}_i, \mathbf{c}_{i^*}(\equiv -\mathbf{c}_i)$. This means that all eigenvectors only have nonzero entries at indices corresponding to $f_i$ or $f_{i^*}$. For the current analysis, it is convenient to write the eigenvalues in polar form:

\begin{equation}
    \lambda_k = r_k e^{j \varphi_k} = r_k \left[\cos\left(\varphi_k\right) + j \sin\left(\varphi_k\right)\right],
\end{equation}
where $r_k = \sqrt{a_k^2 + b_k^2}$ and $\varphi_k=\arctan\left(b_k/a_k\right)$. 

\subsubsection*{Stability and error analysis through eigenvalues}
The amplitude of the eigenvector $\mathbf{v}_k$ is multiplied by $r_k$ at each time step. Therefore, $r_k$ indicates the growth/decay of the eigenvector over time:
\begin{itemize}
    \item $r_k<1$ Decay 
    \item $r_k=1$ Stable
    \item $r_k>1$ Growth
\end{itemize}

For a boundary treatment to be numerically stable, the modes should never grow in time, that is $r_k\leq1$. Furthermore, since streaming represents pure advection, $r_k$ should theoretically be exactly 1. Correspondingly, the diffusive error of the $k$th mode can be expressed as $1-r_k$.

The eigenvector $\mathbf{v}_k$ will oscillate between its real and imaginary values with a frequency $f_k=\varphi_k/(2\pi)$. Given the exact frequency, the dispersive error of the $k$th mode can then be found as the difference with $f_k$. For special cases, the exact frequencies can be found analytically. One such special case is discussed next.


\subsection{2D inclined channel case}\label{subsec:inclinedChannelSetup}
\begin{figure}
    \centering
    \includegraphics[width=0.9\linewidth]{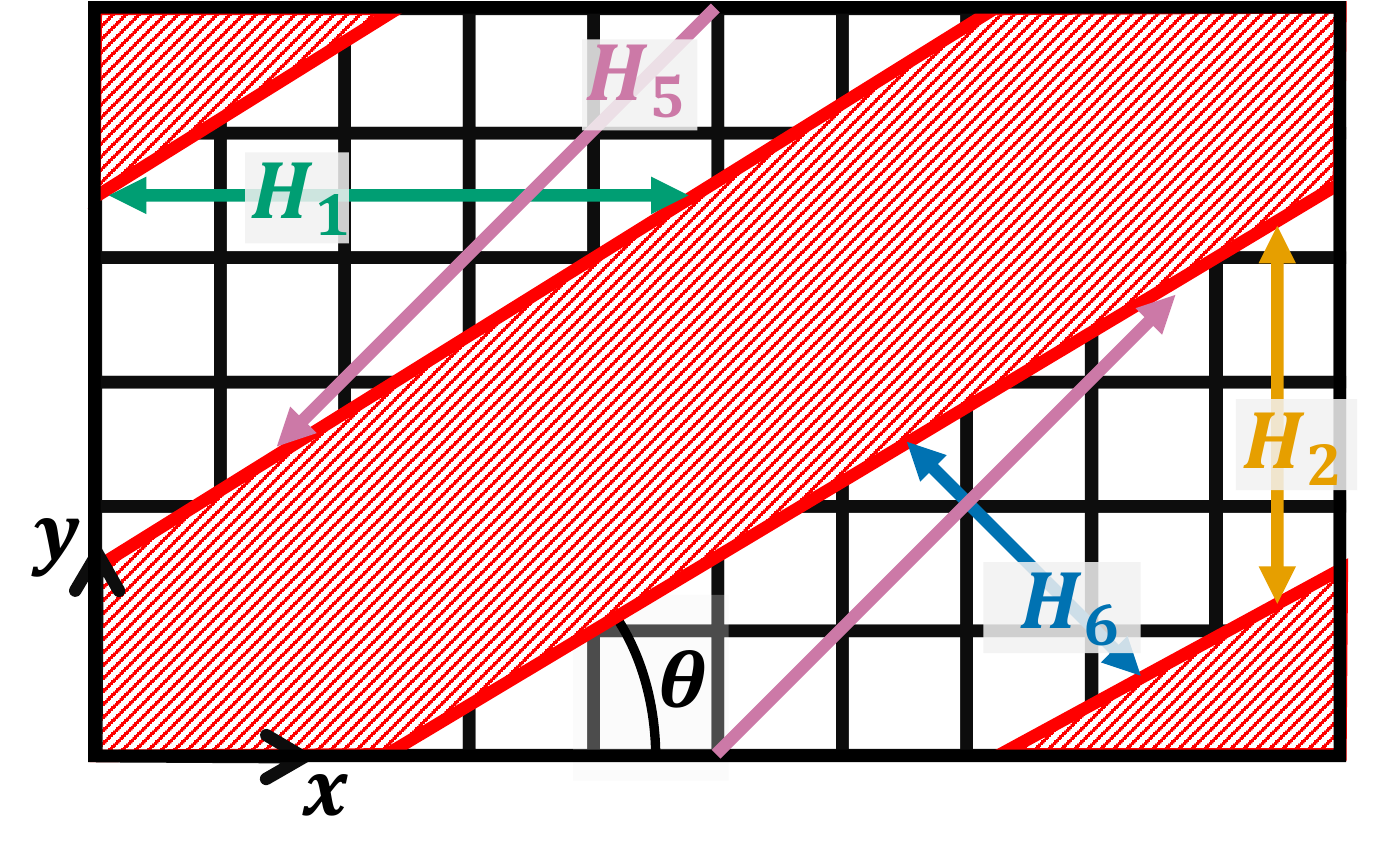}
    \caption{Case setup of the 2D inclined channel. Periodicity is enforced in $x$- and $y$-direction. Arrows indicate the wall separation $H_i$ for each population direction $\mathbf{c}_i$.}
    \label{fig:InclinedChannelCaseSetup}
\end{figure}

To assess the proposed boundary treatment, a 2D case is needed for which: \textbf{1)} the exact frequencies of the eigenvalues are known, and \textbf{2)} the underlying cells are cut at random locations. Both criteria are met by the inclined channel case, which we adapt from Li et al. \citep[p.~43]{Li2012}. The setup of the case is shown in Fig.~\ref{fig:InclinedChannelCaseSetup}. Periodicity is enforced in $x$- and $y$-direction, creating an infinitely long channel with an angle $\theta$ with respect to the lattice. We use the inclination angle $\theta=30^\circ$, such that the second criterion is satisfied (cells cut at random locations). Satisfaction of the first criterion is demonstrated below.

\subsubsection*{Analytical eigenvalue frequencies in 1D}
\begin{figure}
    \centering
    \includegraphics[width=\linewidth]{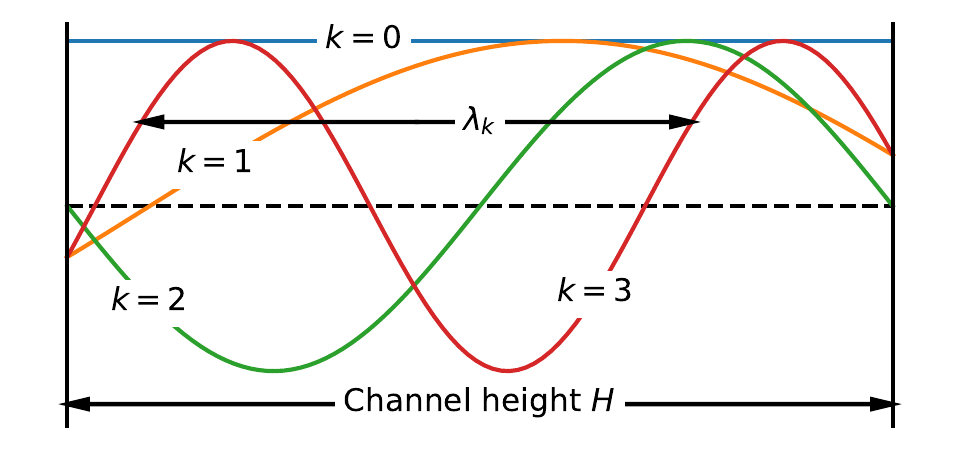}
    \caption{Analytical eigenvectors for $k=0,1,2,3$ in a 1D domain bounded by two walls. Only their component in $f_i$ is shown.}
    \label{fig:eigenvectors1D}
\end{figure}

We first derive the analytical eigenvalue frequencies for a 1D domain bounded by two walls, separated by a height $H$. The analytical eigenvectors comprise waves traveling between the walls with wavelengths $\lambda_k$. These live in both $f_i$ and $f_{i^*}$, traveling with the corresponding velocity $\mathbf{c}_i$ or $\mathbf{c}_{i^*}$($\equiv-\mathbf{c}_i$). For each wavelength, there are two eigenvectors which are complex conjugates. An example of the $f_i$ component of these eigenvectors is shown for $k=0,1,2,3$ in Fig.~\ref{fig:eigenvectors1D}. The component in $f_{i^*}$ looks similar, only phase shifted to connect smoothly at the boundaries.

The analytical equation for the wavelength $\lambda_k$ follows directly from Fig.~\ref{fig:eigenvectors1D}. For the frequency $f_k$, with wave speed $c=f_k \lambda_k$:
\begin{equation}
    \lambda_k = 2H/k; \qquad f_k = c/\lambda_k =  \frac{\Delta x}{\Delta t}\frac{1}{\lambda_k},
\end{equation}
where $\Delta t$ is the timestep and $\Delta x$ is the grid size. In standard streaming, populations travel one grid distance per timestep, so that $c = \Delta x / \Delta t$.

\subsubsection*{Analytical eigenvalue frequencies in 2D}
We now explain how to obtain the analytical eigenvalue frequencies for the 2D inclined channel (Fig.~\ref{fig:InclinedChannelCaseSetup}). Consider again the distance between two walls, but now along the direction of each opposite-moving population pair $\mathbf{c}_i$/$\mathbf{c}_{i^*}$. This distance is denoted by $H_i$, and is constant along the channel. The distances $H_{i}$ are indicated in Fig.~\ref{fig:InclinedChannelCaseSetup}. Along a cut in this direction, the analytical eigenvectors will again comprise waves traveling between the channel walls, living in both $f_i$ and $f_{i^*}$. Their wavelengths $\lambda_{k,i}$ and frequencies $f_{k,i}$ are:
\begin{equation}
    \lambda_{k,i}=2H_i/k; \qquad f_{k,i} = c/\lambda_{k,i} = \frac{\left|\mathbf{c}_i\right|}{\lambda_{k,i}}\frac{\Delta x}{\Delta t}.
\label{eq:2DEigenFrequencies}
\end{equation}

Where the wave speed is $c=\left|\mathbf{c}_i\right| \Delta x / \Delta t$, since diagonal populations travel a distance $\sqrt{2}\Delta x$ during one timestep. Because the channel is infinitely long, the analytical eigenvectors will be the same irrespective of the cut location. 

It should be mentioned that in following results for the discretized system, there will be multiple eigenvalues per harmonic. To explain this, consider an eigenmode of horizontal populations, representing a wave traveling in horizontal direction. This eigenmode can have any shape in the vertical direction. Thus, the number of eigenmodes with this frequency will be equal to twice the number of cells in the vertical direction, taking also the complex conjugates into account. Similarly, eigenmodes of vertical and diagonal populations have a given multiplicity depending on the domain size and channel angle. This will be visible in the results. 


\subsection{Eigenvalue analysis}\label{subsec:eigenvalueResults}
The method to obtain the eigenvalues of the streaming matrix, explained in Sec.~\ref{subsec:conceptEigenValues}, is now applied to the inclined channel laid out in Sec.~\ref{subsec:inclinedChannelSetup}. We use the proposed boundary treatment which was explained in Sec.~\ref{sec:methodology}. 

\subsubsection*{Stability and diffusive error analysis}
\begin{figure*}
    \centering
    \includegraphics[width=\linewidth]{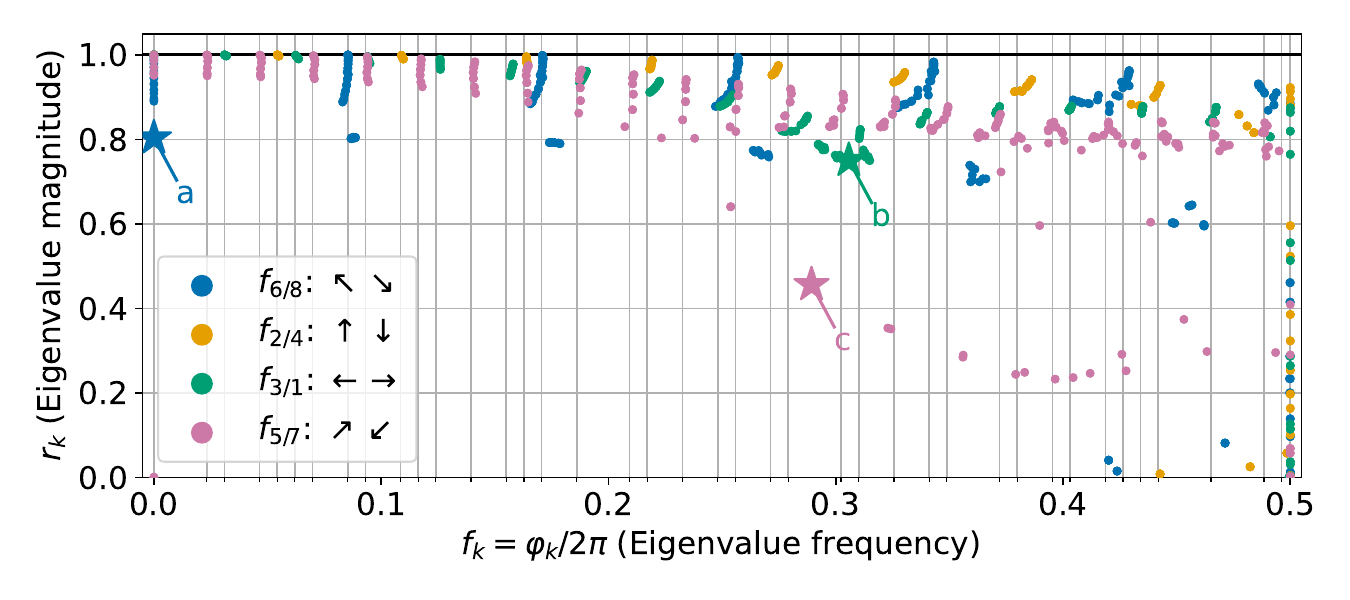}
    \caption{Magnitude versus frequency, for the eigenvalues of the streaming matrix. The geometry used is the inclined channel (see Sec.~\ref{subsec:inclinedChannelSetup}) at inclination angle $\theta=30^\circ$ and with a channel width of $8\Delta t$. Theoretical harmonics are indicated on the $x$-axis. The eigenmodes corresponding to the eigenvalues marked a-c are shown in Fig.~\ref{fig:eigenmodes}.}
    \label{fig:eigsMagnitude}
\end{figure*}

In order to assess stability, the $r_k$ values are plotted against $f_k$ in Fig.~\ref{fig:eigsMagnitude}, where the theoretical frequencies are indicated on the $x$-axis. Specifically, the inclined channel with inclination angle $\theta=30^\circ$ and a channel width of $8\Delta x$ is used. The stability condition $r_k\leq 1$ is verified to hold within machine precision. Thus, it is concluded that the proposed boundary treatment is stable. 

\begin{figure*}
    \begin{subfigure}[b]{0.32\textwidth}
        \includegraphics[width=\linewidth]{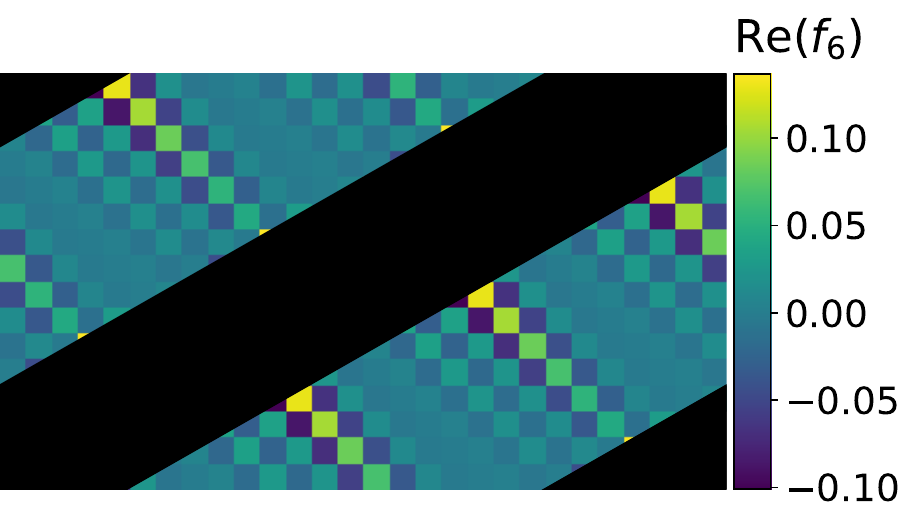}
        \caption{}
        \label{eigenmodes_a}
    \end{subfigure}
    \begin{subfigure}[b]{0.32\textwidth}
        \includegraphics[width=\linewidth]{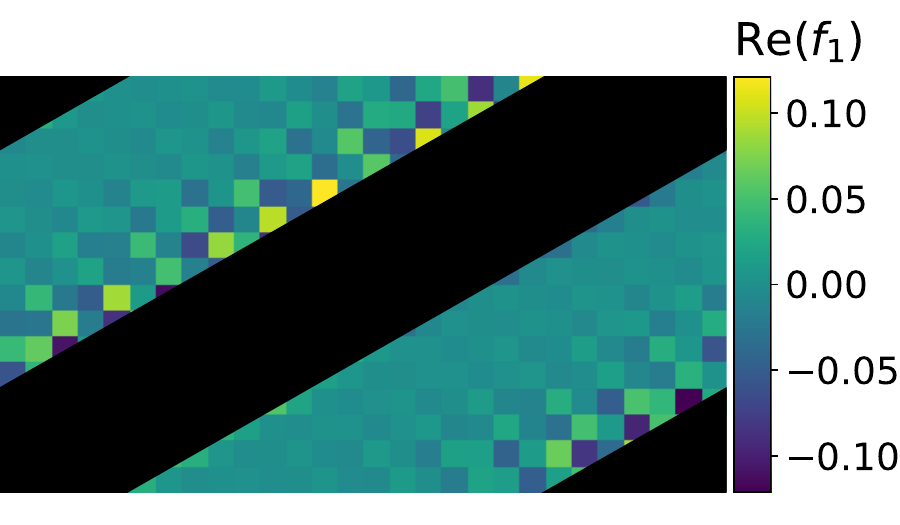}
        \caption{}
        \label{eigenmodes_b}
    \end{subfigure}
    \begin{subfigure}[b]{0.32\textwidth}
        \includegraphics[width=\linewidth]{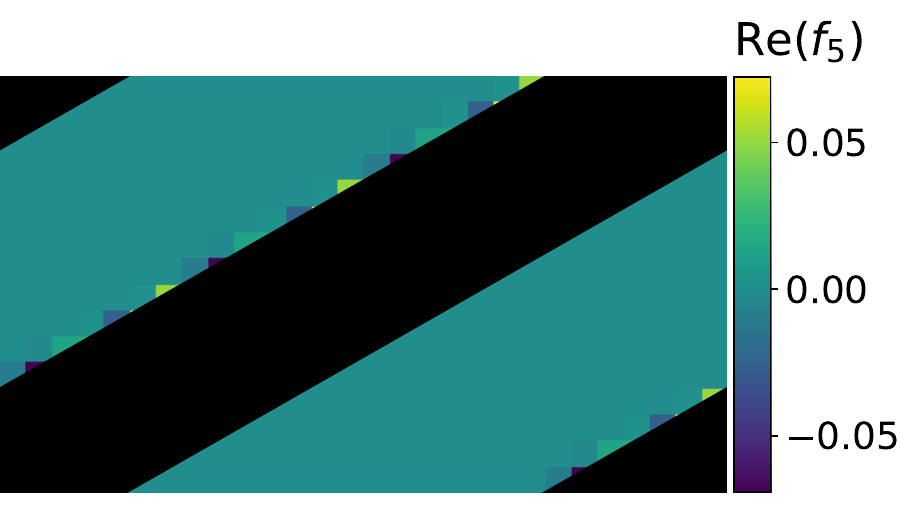}
        \caption{}
        \label{eigenmodes_c}
    \end{subfigure}
    \caption{Eigenmodes corresponding to the eigenvalues marked in Fig.~\ref{fig:eigsMagnitude}. The real component of one of the two populations is shown.}
    \label{fig:eigenmodes}
\end{figure*}

Note that the observed stability is not coincidental. During the design of the boundary treatment, we evaluated this stability condition for various algorithmic choices. Two were identified that make the boundary treatment unstable:
\begin{itemize}
    \item The boundary-weighted method resulted in $r_k>1$ (see Sec.~\ref{subsubsec:PP1D}).
    \item Including a bilinear contribution ($xy$ term) in the 2D discontinuous piecewise linears resulted in $r_k>1$ (see Sec.~\ref{subsubsec:PP2D}).
\end{itemize}

In Fig.~\ref{fig:eigsMagnitude} it is observed that $r_k$ generally decreases as $f_k$ increases. However, even at lower frequencies there are some eigenvalues with a significantly lower $r_k$. To further investigate, we show the eigenmodes associated with three eigenvalues (marked a-c in Fig.~\ref{fig:eigsMagnitude}) in Fig.~\ref{fig:eigenmodes}. Modes a exhibits oscillations with the smallest possible wavelength in the direction perpendicular to $\mathbf{c}_i$, meaning it is not well-resolved on the grid and thus grid-dependent. The different tracks in this mode mix at the boundaries in the direction perpendicular to $\mathbf{c}_i$. Modes b and c live only along the boundary and thus come purely from the boundary treatment, also making them nonphysical. The mixing between adjacent voxels in the construction of the piecewise linears and geometric mapping dissipates these modes. In typical problems, these nonphysical and grid-dependent modes are rarely excited. Thus, the fact that they are damped out quickly does not strongly affect accuracy, but may further enhance stability.

\subsubsection*{Dispersive error analysis}
\begin{figure*}
    \begin{subfigure}[b]{0.49\textwidth}
        \includegraphics[width=\linewidth]{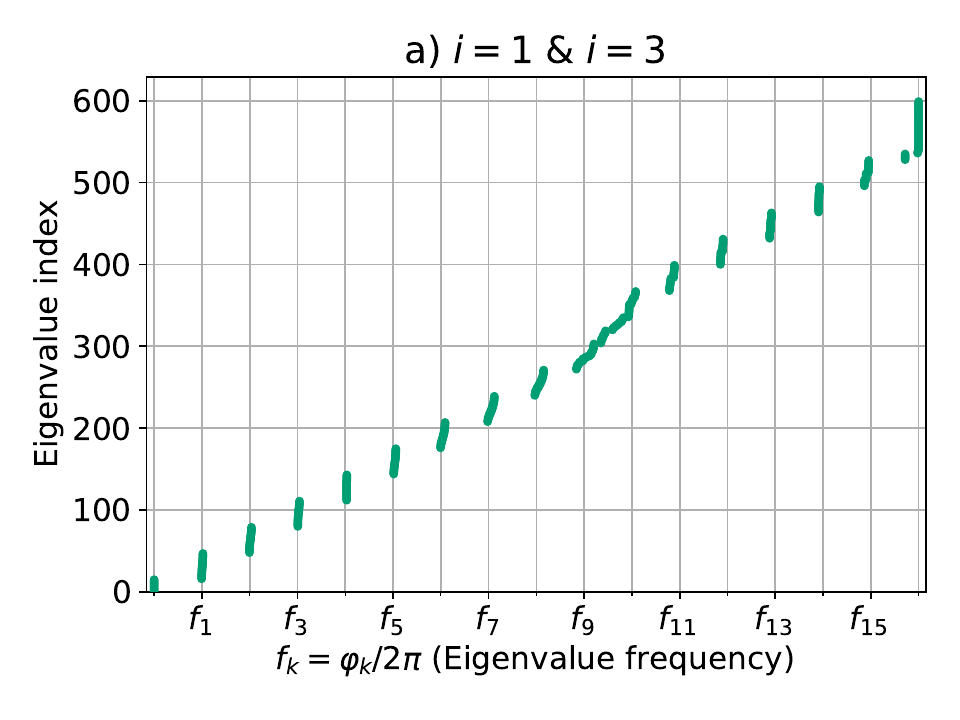}
    \end{subfigure}
    \begin{subfigure}[b]{0.49\textwidth}
        \includegraphics[width=\linewidth]{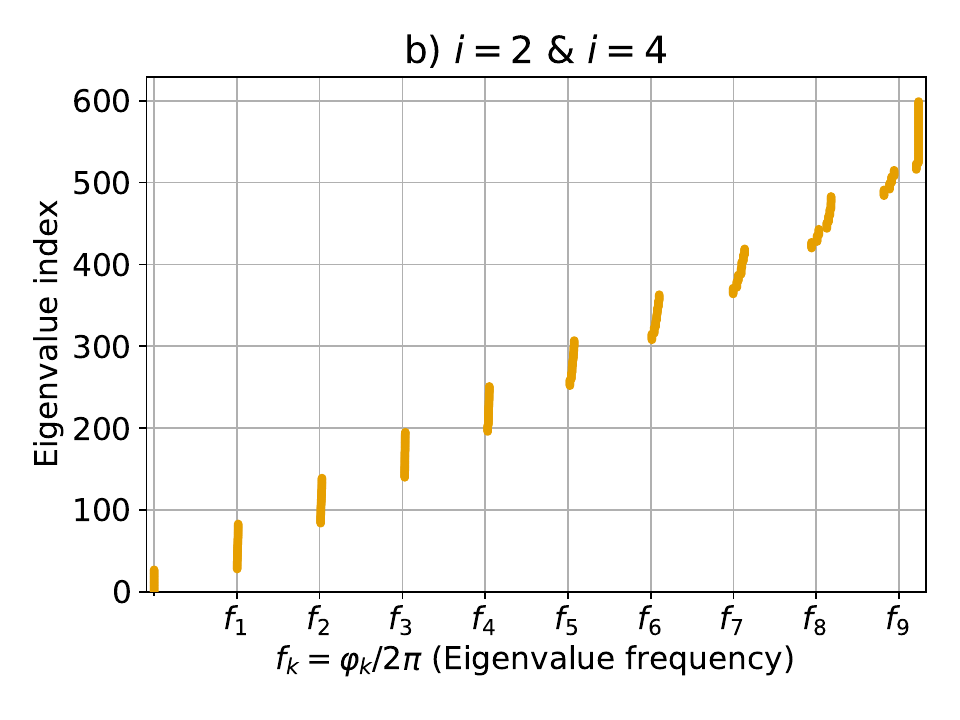}
    \end{subfigure}
    \begin{subfigure}[b]{0.49\textwidth}
        \includegraphics[width=\linewidth]{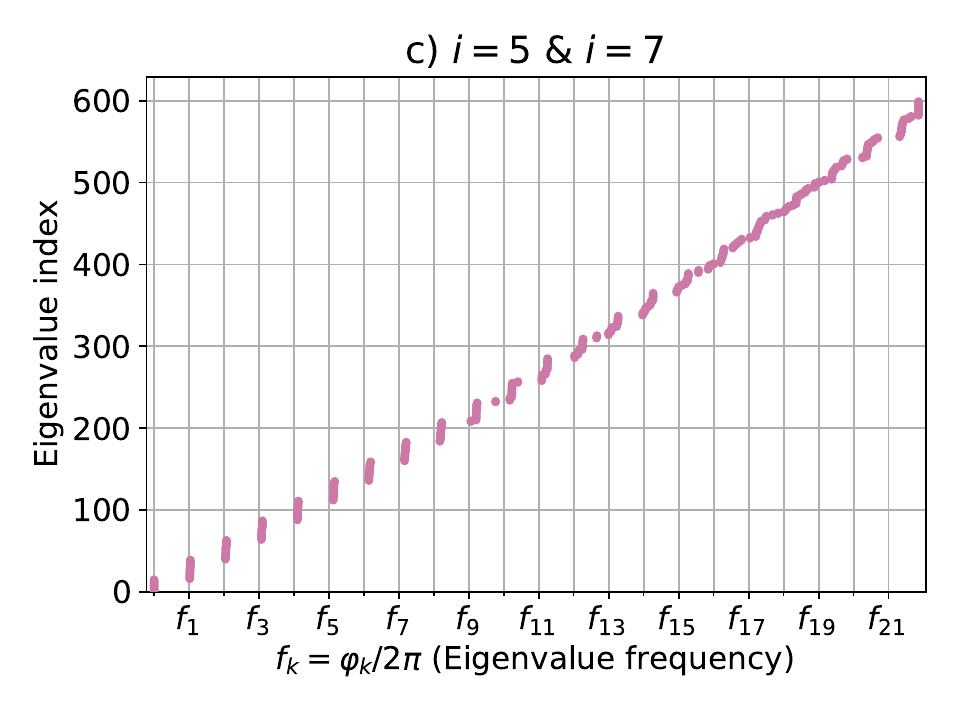}
    \end{subfigure}
    \begin{subfigure}[b]{0.49\textwidth}
        \includegraphics[width=\linewidth]{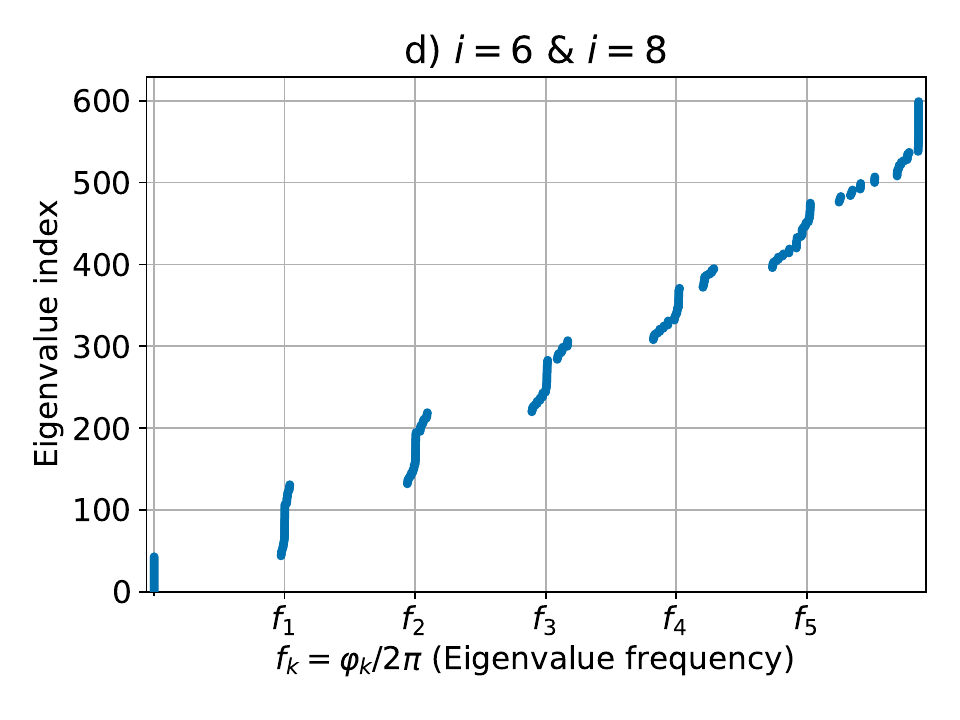}
    \end{subfigure}
    \caption{Sorted eigenvalue index versus its frequency. The geometry used is the inclined channel (see Sec.~\ref{subsec:inclinedChannelSetup}) at inclination angle $\theta=30^\circ$ and with a channel width of $8\Delta t$. Theoretical harmonics are indicated on the $x$-axis.}
    \label{fig:sortedEigs}
\end{figure*}

For the following analysis, we consider each opposite-moving population pair separately. The corresponding eigenvalues are sorted by frequency and plotted in Fig.~\ref{fig:sortedEigs}. The theoretical base frequency $f_1$ and its higher harmonics $f_k$ (based on Eq.~\ref{eq:2DEigenFrequencies}) are indicated on the $x$-axis. 

We observe that eigenvalues cluster near one of the theoretical harmonics $f_k$. The distance between the eigenvalue's frequency and the theoretical one directly relates to the dispersive error. For higher frequencies, the dispersive error increases. This is especially noticeable in Fig.~\ref{fig:sortedEigs}d. As previously discussed, these high-frequency modes are not captured well by the grid and strongly damped, limiting their overall impact on the solution.

\section{Numerical results}\label{sec:ConvergenceResults}
In order to assess the accuracy of the proposed boundary treatment, we applied it to three test cases: \textbf{1)} a quasi-1D, gravity-driven channel, \textbf{2)} a 2D, gravity-driven channel which is inclined relative to the grid, and \textbf{3)} flow around a NACA0012 airfoil. Cases 1 and 2 were used to establish the mathematical properties of the method, such as order of convergence. Case 3 was used to demonstrate the method in a more practical setting. Cases 1 and 2 are taken from Li. \cite{Li2012} for which we simulated the same datapoints. In that work, Li compares two existing boundary algorithms: one proposed by Chen et al. \cite{Chen1998b} and one by Li et al. \cite{Li2009}. We implemented both algorithms in order to benchmark our method. Our implementation of Chen et al.'s algorithm produced results consistent with those reported in \cite{Li2012}. However, we were unable to reproduce the results of Li et al.'s algorithm with our own implementation. Therefore, in our plots, we included only results from our implementation of Chen et al.'s algorithm. Nevertheless, we do report convergence rates of Li et al. These were computed from the data in their published figures and reported as `Li et al. (digitized)'. 

Finally, although we present only the $L_1$ error norms here, we also tested the $L_2$ and $L_\infty$ error norms, any differences are discussed.

\subsection{Velocity convergence in a quasi-1D channel}\label{subsec:velConv1D}
\begin{figure}
    \centering
    \includegraphics[width=\linewidth]{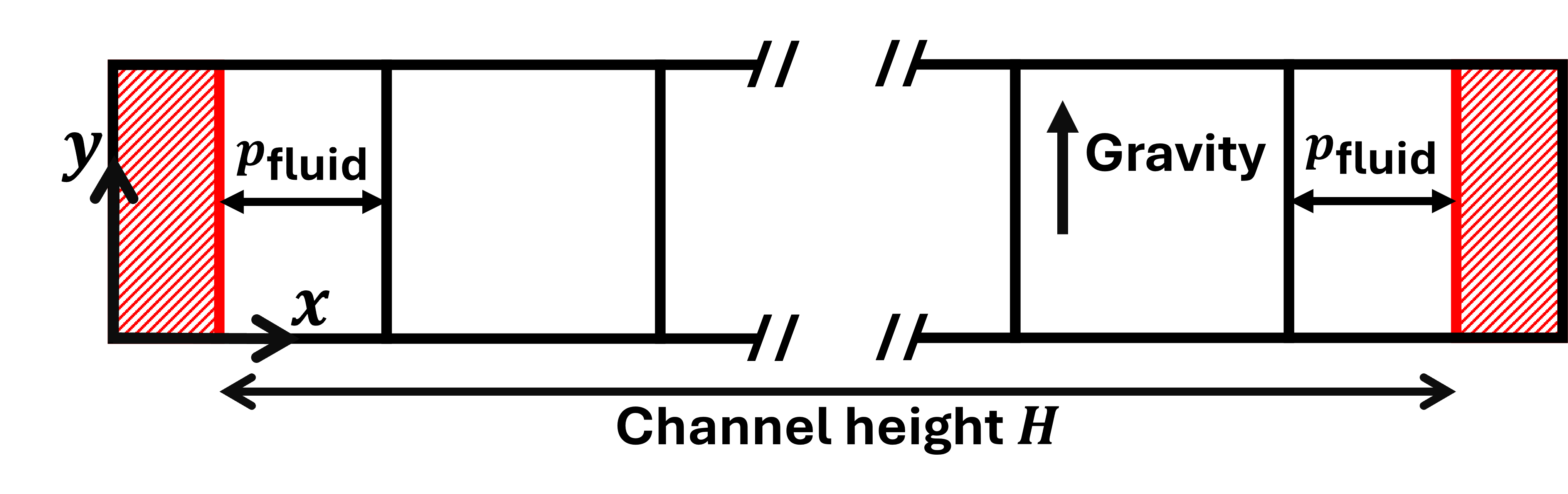}
    \caption{Case setup of the quasi-1D, gravity-driven channel flow. Periodicity is enforced in $y$-direction.}
    \label{fig:Q1DGravityChannelCaseSetup}
\end{figure}

To test convergence of the velocity in 1D, we used the quasi-1D channel presented by Li \citep[p.43]{Li2012}. The setup is shown in Fig.~\ref{fig:Q1DGravityChannelCaseSetup}. No-slip was enforced at the left and right walls, while periodicity was enforced in $y$-direction. A gravity force was applied in $y$-direction. This resulted in a parabolic $y$-velocity profile for which the exact solution is known \cite{Li2012}:
\begin{equation}
    U = -\frac{g}{2\nu}x^2 + \frac{g}{2\nu}H x,
\label{eq:UGravityChannel}
\end{equation}
where $H$ is the channel height and $g$ is the gravitational constant. The left and right walls were intentionally offset with respect to the lattice to create a cut cell at each wall. These cut cells are thus part solid and part fluid, as shown in Fig.~\ref{fig:Q1DGravityChannelCaseSetup}. The fluid fraction of these cut cells, $p_{\text{fluid}}$, was varied in the subsequent study. 

\begin{figure}
    \centering
    \includegraphics[width=\linewidth]{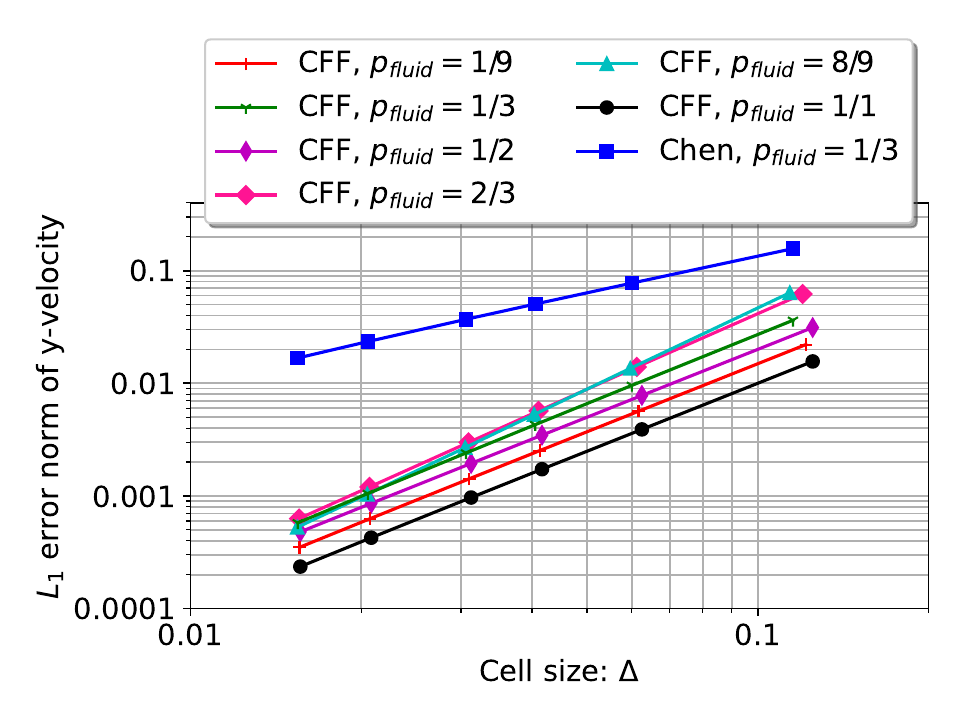}
    \caption{Mesh convergence study of the quasi-1D gravity-driven channel with the volume-weighted CFF algorithm for various $p_{\text{fluid}}$ and the algorithm of Chen et al. \cite{Chen1998b} for $p_{\text{fluid}}=1/3$.}
    \label{fig:Q1DGC}
\end{figure}

The mesh convergence study was performed by refining the grid such that cell size $\Delta$ decreased. When refining, the boundaries were slightly shifted to retain the same $p_{\text{fluid}}$ value. Refinement implies an increase in the non-dimensional relaxation time $\tau^*$. We made sure to keep $\tau^*$ below 0.6 to limit the slip velocity error \cite[p.~188]{Kruger2017}. The convergence study was performed for various $p_{\text{fluid}}$ for the volume-weighted CFF algorithm as well as the original volumetric scheme by Chen et al. \cite{Chen1998b}. The $L_1$ error norm of the velocity is shown in Fig.~\ref{fig:Q1DGC}, with results from Chen et al. only shown for $p_{\text{fluid}}=1/3$. Also, the order of convergence values obtained from a least-squares fit to the data are reported in Tab.~\ref{tab:Q1DGC}.

\begin{table}
    \centering
    \caption{Comparison of the order of convergence for the quasi-1D gravity-driven channel at various $p_{\text{fluid}}$.}
    \label{tab:Q1DGC}
    \begin{tabular}{|c|c|c|c|c|c|c|} \hline
    & \multicolumn{6}{c|}{$p_{\text{fluid}}$} \\ \hline
    & 1/9 & 1/3 & 1/2 & 2/3 & 8/9 & 1/1 \\ \hline \hline
    \textbf{Chen} & 1.16 & 1.11 & 1.11 & 1.15 & 1.33 & 2.02\\ \hline
    \textbf{Li} (digitized) & 1.25 & 1.73 & 1.70 & 1.80 & 1.89 & 2.06 \\ \hline
    \textbf{CFF} & 2.02 & 2.06 & 2.01 & 2.25 & 2.40 & 2.02 \\ \hline
    \end{tabular}
\end{table}

All methods obtain an order of convergence of $\sim2$ for $p_{\text{fluid}}=1/1$. This case corresponds to halfway bounce-back and the order of 2 is widely reported in literature \cite{Zou1997}. Next, for small $p_{\text{fluid}}$, CFF exhibits an order of around 2, while Chen and Li are closer to 1. For larger $p_{\text{fluid}}$, Li is also close to 2, while CFF is even a bit higher than 2 for coarse grids. Though CFF approaches order 2 for finer grids. In conclusion, CFF exhibits at least second-order accuracy across the full range of $p_{\text{fluid}}$.


\subsection{Population convergence in 1D}\label{subsec:speciesConv1D}
In the previous section, the convergence of the velocity was studied. The final velocity field depends on three aspects of the algorithm: streaming, collision, and the implementation of the body force. In order to test the boundary treatment in an isolated setting, this section considers a case that uses only the streaming step. This corresponds to pure advection of the populations with their own velocity. A 1D domain with two no-slip walls is used.

\begin{figure}
    \centering
    \includegraphics[width=\linewidth]{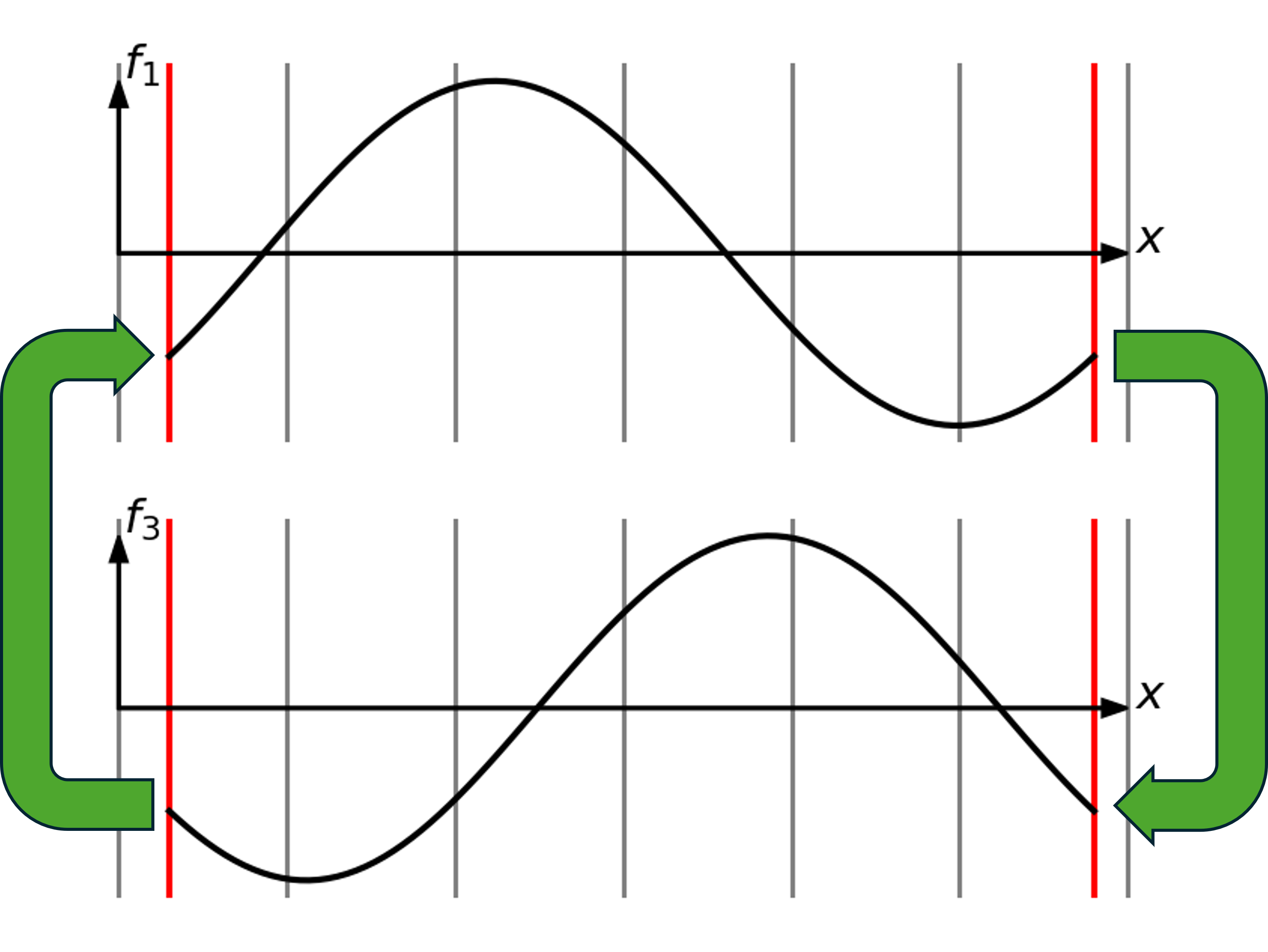}
    \caption{Illustration of cycling between opposite-moving populations in the 1D pure advection case.}
    \label{fig:cyclingSine}
\end{figure}

For this advection case, each population must be initialized as a continuous, smoothly varying field. In the bulk, this field theoretically shifts a distance $\mathbf{c}_i$/$\mathbf{c}_{i^*}$ at each time step. This is exactly satisfied by the bulk streaming operation in the LBM. Thus, any observed errors can be attributed to the boundary treatment. When the field of population $f_i$ streams into a no-slip boundary, its velocity is reversed, such that it streams out as the opposite population $f_{i^*}$. As a result, the field appears to cycle between two no-slip boundaries, switching to the opposite population at each one. This cycling is illustrated in Fig.~\ref{fig:cyclingSine}. The connection between $f_i$ and $f_{i^*}$ at the boundary implies that their fields must connect smoothly there at initialization.


This pure advection case is now considered for a 1D channel. The following exact solution was used for both the left-moving and right-moving populations:
\begin{equation}
    f_i(x, t) = \frac{1}{2} - \frac{1}{2}\cos\left(\frac{2\pi}{H}\left(x - x_l - c_i t\right)\right),
\end{equation}
where $H$ is the channel height and $x_l$ is the location of the left boundary. This function ensures that at $t=0$, $f_i$ and its derivative are zero at the boundaries, satisfying the smoothness requirements. Of course, gradients will be nonzero at the boundaries in further timesteps.

\begin{figure}
    \centering
    \includegraphics[width=\linewidth]{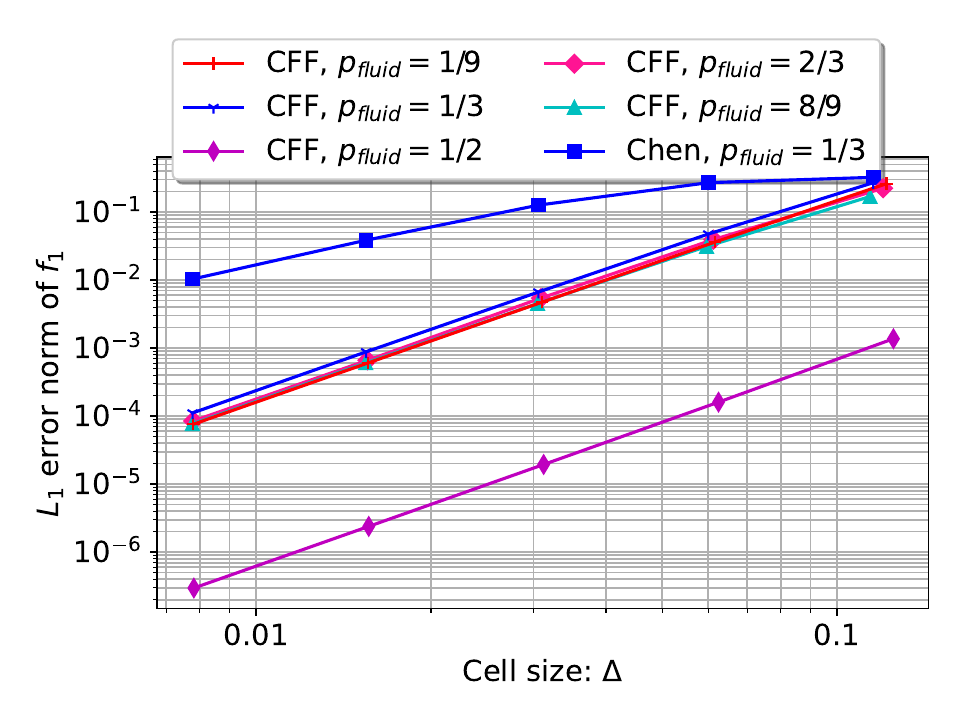}
    \caption{Mesh convergence study of pure advection in 1D. The $L_1$ error norm of the right-moving population ($f_1$) is plotted against cell size. Two algorithms are shown: volume-weighted CFF for various $p_{\text{fluid}}$ and the algorithm of Chen et al. \cite{Chen1998b} for $p_{\text{fluid}}=1/3$.}
    \label{fig:streamingOnly1D}
\end{figure}

The case was run at various grid refinements. The coarsest case, corresponding to $\sim8\Delta x$ separation between boundaries, was run for 501 time steps. The number of time steps was increased proportionally for refined cases, so that e.g. 1002 time steps were used with $\sim16\Delta x$ separation between boundaries. This case allows the study of the convergence of individual populations. This was not done previously, since an exact solution was only available for velocity. The error of population $f_1$ is shown for various $p_{\text{fluid}}$ in Fig.~\ref{fig:streamingOnly1D}. For all $p_{\text{fluid}}$, CFF converges with approximately third order in $L_1$. Meanwhile, the algorithm of Chen et al. converges with approximately second order in $L_1$. Thus, both methods converge with one order higher accuracy in $L_1$ compared to the case with collision in Sec.~\ref{subsec:velConv1D}.

Interestingly, we found different convergence orders for the other norms: for the $L_2$ norm, CFF and Chen converged with order 2.5 and 1.5, respectively; for the $L_\infty$ norm with order 2 and 1, respectively. To investigate this observation, we analyze the convergence of CFF by deriving an analytical expression for the error as a function of the cell size $\Delta$, using the symbolic manipulation package Sympy \cite{sympy}. This is only done for CFF.

We find that the error in the cut cells at the edges of the domain is second order in cell size ($\mathcal{O}(\Delta^2)$), which explains the convergence of the $L_\infty$ norm. After an initialization effect, however, a third-order convergence ($\mathcal{O}(\Delta^3)$) is observed in the cells adjacent to the cut cells (cell 1 in Fig.~\ref{fig:1DMapping_zones}), which propagates to the bulk. This property holds for all $p_{fluid}$, explaining the observed convergence rate of 3 in the $L_1$ norm and rate 2.5 in the $L_2$ norm. We hypothesize that this is a fundamental property of mass conserving schemes since we observe the same order increase for the scheme of Chen et al.




Note that the order increase only occurs with pure advection. Thus, it does not occur for cases with collision, where populations are redistributed at each timestep. This means that the complete method with CFF boundary treatment is still a second-order method (and Chen a first-order one), which is consistent with our observations in Sec.~\ref{subsec:velConv1D}.

\subsection{Velocity convergence in a 2D channel}\label{subsec:velConv2D}
To test the convergence of velocity in 2D, we used the inclined channel presented by Li et al. \cite[p.~43]{Li2012}, as described in Sec.~\ref{subsec:inclinedChannelSetup}. To drive the flow, a gravity force was applied in the longitudinal direction of the channel, which theoretically results in a parabolic velocity profile in the transverse direction. This profile is described by Eq.~\ref{eq:UGravityChannel}, where $x$ now represents the distance to the wall.

\begin{figure}
    \centering
    \includegraphics[width=\linewidth]{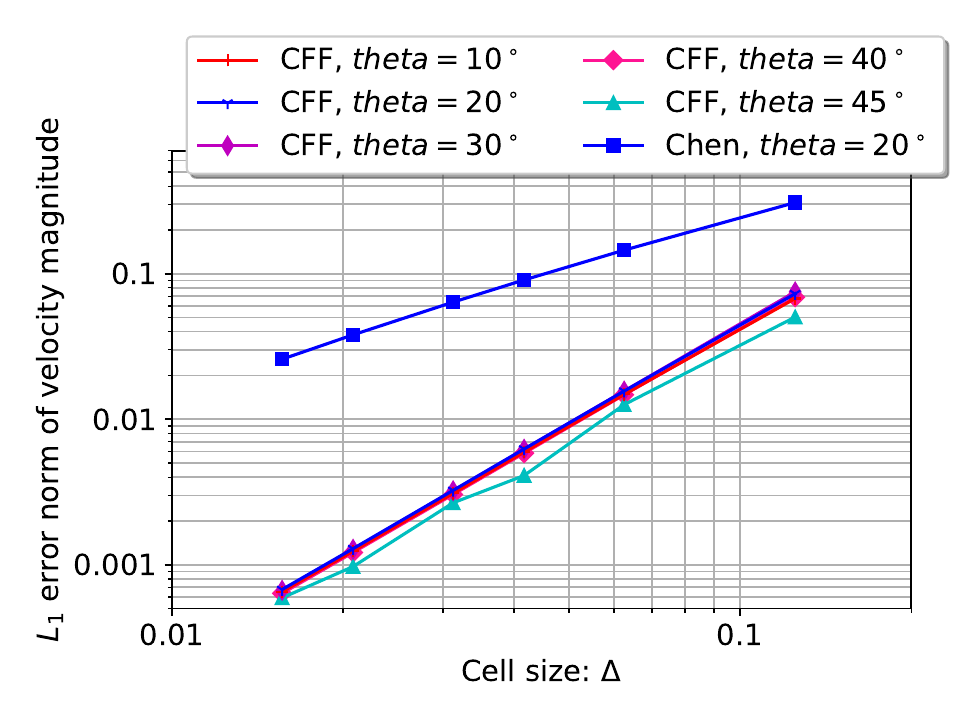}
    \caption{Mesh convergence study of the inclined gravity-driven channel with the volume-weighted CFF algorithm for various inclination angles $\theta$ and the algorithm of Chen et al. \cite{Chen1998b} for $\theta=20^\circ$.}
    \label{fig:IC2D}
\end{figure}

A mesh convergence study was performed in which $\tau^*$ was kept below 0.6. The study was performed for various channel inclination angles $\theta$ for CFF as well as the original volumetric scheme by Chen et al. \cite{Chen1998b}. The results are shown in fig.~\ref{fig:IC2D}, with results from Chen et al. \cite{Chen1998b} only shown for $\theta=20^\circ$. The order of convergence was again obtained with a least-squares fit and is reported in Tab.~\ref{tab:IC2D}. For the order of convergence, Chen lies around 1.15, Li around 1.6, and CFF around 2.2. In terms of the absolute error, CFF also performs better than both Chen and Li. 

Finally, we consider another measure of accuracy: the isotropy, which reflects the dependence on the orientation of the lattice. In Fig.~\ref{fig:IC2D}, the error curves for different $\theta$ align closely, indicating that CFF behaves isotropically, as desired. In contrast, Li reports significant variation with $\theta$ \citep[p.~54]{Li2012}. This anisotropic behaviour is also observed by Goyal et al. \cite{Goyal2024} in the commercial solver PowerFLOW, which implements a variation of Li.

\begin{table}
    \centering
    \caption{Comparison of the order of convergence for the inclined gravity-driven channel at various inclination angles $\theta$.}
    \label{tab:IC2D}
    \begin{tabular}{|c|c|c|c|c|c|} \hline
    & \multicolumn{5}{c|}{Inclination angle $\theta$} \\ \hline
    & $10^\circ$ & $20^\circ$ & $30^\circ$ & $40^\circ$ & $45^\circ$ \\ \hline \hline
    \textbf{Chen} & 1.19 & 1.19 & 1.21 & 1.22 & 1.17 \\ \hline
    \textbf{Li} (digitized) & 1.46 & 1.49 & 1.56 & 1.71 & 1.60  \\ \hline
    \textbf{CFF} & 2.24 & 2.25 & 2.27 & 2.26 & 2.17 \\ \hline
    \end{tabular}
\end{table}

\subsection{Population convergence in a 2D channel}\label{subsec:speciesConv2D}
The pure advection approach outlined in Sec.~\ref{subsec:speciesConv1D} is now applied to the 2D inclined channel geometry. The field is initialized as:

\begin{eqnarray}
    f_i(x, y, t=0) = &&\left[ \frac{1}{2} - \frac{1}{2}\cos\left(\frac{2\pi}{H_x}\left\{x - x_l(y)\right\}\right)\right] \cdot \\ \nonumber
    &&\cdot\left[ \frac{1}{2} - \frac{1}{2}\cos\left(\frac{2\pi}{h_y}y\right)\right],
\end{eqnarray}
where $H_x$ is the channel height in $x$-direction and $h_y$ is the domain height (not channel height) in $y$-direction. Furthermore, $x_l$ is the $x$-coordinate of the left boundary of the channel, which depends on the $y$-coordinate (see Fig.~\ref{fig:InclinedChannelCaseSetup}). Since $f_i$ and its derivatives are zero on the boundary at $t=0$, the smoothness constraints described before are satisfied. Of course, gradients will be nonzero at the boundaries in further timesteps.

\begin{figure}
    \centering
    \includegraphics[width=\linewidth]{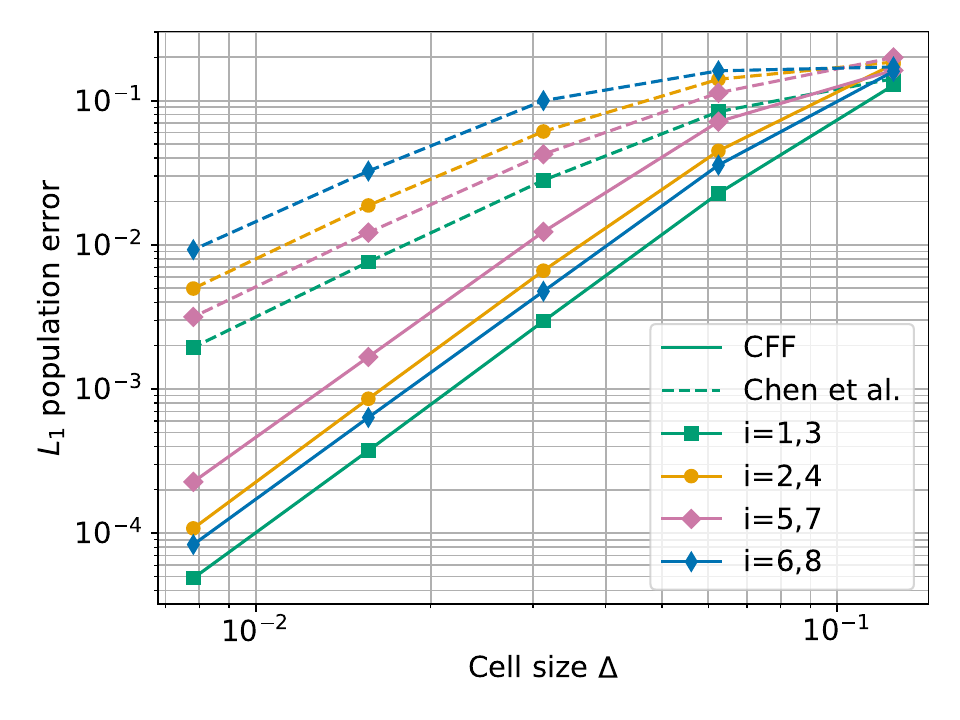}
    \caption{Mesh convergence study for the inclined channel with pure advection. The $L_1$ error norm of each opposite-moving population pair is plotted against cell size. A channel inclination angle $\theta=30^\circ$ is used.}
    \label{fig:cyclingSineConvergence}
\end{figure}

The case was run at various grid refinements. For the coarsest case, with $\sim8\Delta x$ separation between boundaries, 501 time steps were used. The number of timesteps was increased proportionally for refined cases. We again consider the error of the populations directly. In 2D, this results in four errors; one for each opposite-moving population pair. The results for the CFF and our implementation of Chen et al. are shown in Fig.~\ref{fig:cyclingSineConvergence}, for an inclination angle of $\theta=30^\circ$. For each population pair, we observe a convergence rate of approximately 3 for CFF and approximately 2 for Chen in the $L_1$ norm. For the $L_2$ norm we observe 2.5 and 1.5, respectively and for the $L_\infty$ norm 2 and 1, respectively. This is in line with the observations for 1D.

\subsection{2D Flow around a NACA0012 airfoil}\label{subsec:airfoilResults}
\begin{figure}
    \centering
    \includegraphics[width=\linewidth]{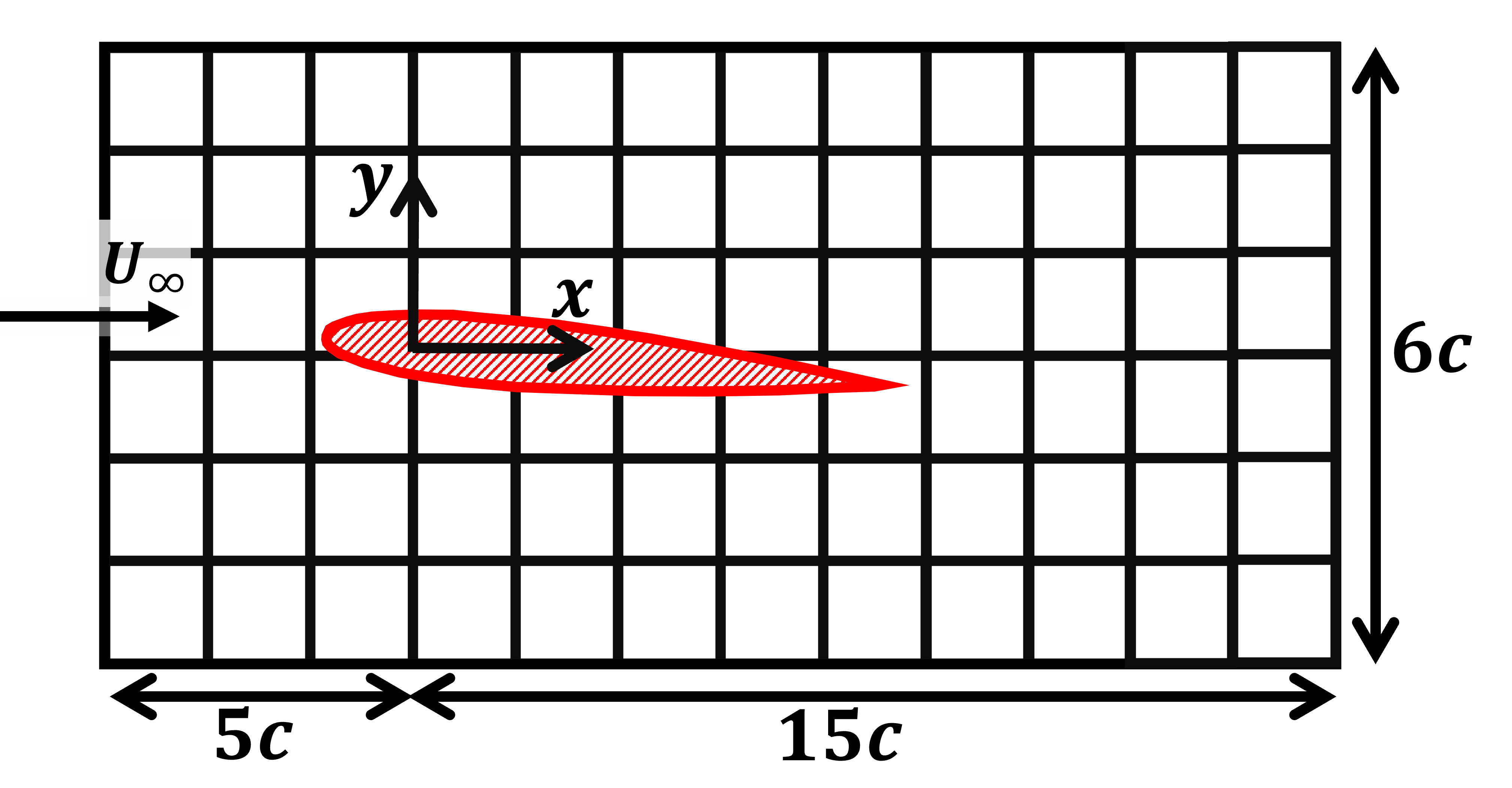}
    \caption{Case setup of the 2D NACA0012 airfoil case (mesh not to scale).}
    \label{fig:2DNacaCase}
\end{figure}
To move closer to a practical application, the 2D laminar flow around the NACA0012 airfoil was simulated, for a 5 degree angle of attack and a chord Reynolds number $Re_c=U_\infty c / \nu$ of 100. The domain height was restricted, resulting effectively in the simulation of an airfoil in a closed wind tunnel. This domain restriction allowed for a uniform mesh, to overcome complications induced by local refinements. The schematic of the case is shown in Fig.~\ref{fig:2DNacaCase}. Boundary conditions were applied as follows: pressure and velocity at the inlet, zero gradient at the outlet, and free-slip at the top and bottom. All cases were run until a steady state was reached.

In order to obtain a ground truth solution, the same case was also run on a fine, body-fitted mesh in the finite-volume solver OpenFOAM \cite{OpenFOAMv7}. The simpleFOAM solver was used. Furthermore, the boundary conditions were slightly adapted to a velocity inlet and a pressure outlet. We also verified mesh independence of the OpenFOAM results.

\begin{figure*}
    \begin{subfigure}[b]{0.49\textwidth}
        \centering   
        \includegraphics[width=\linewidth]{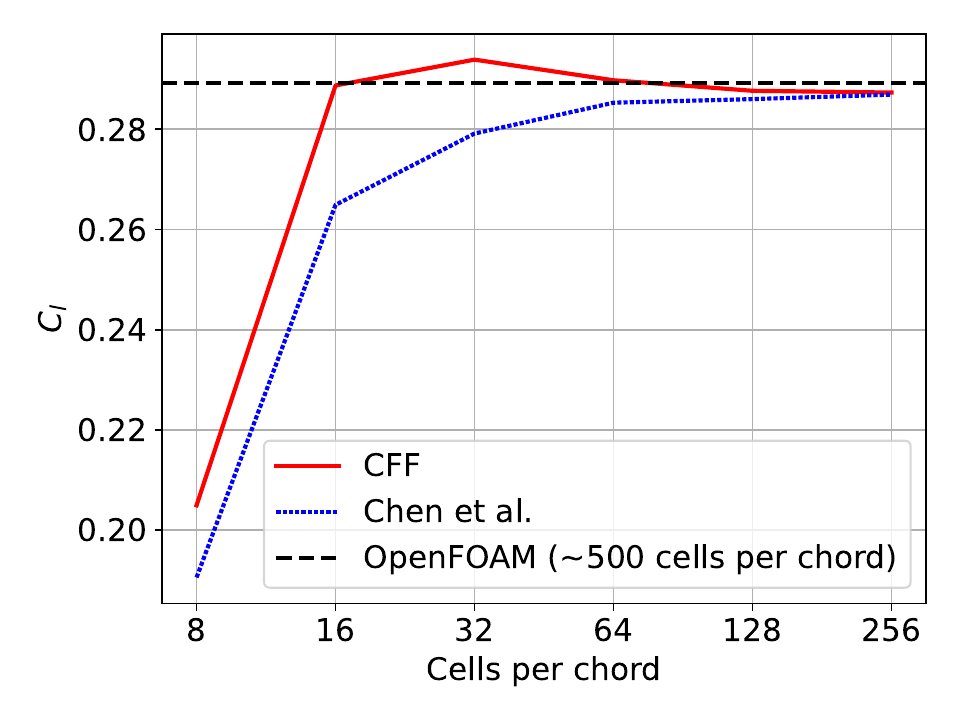}
        \caption{Lift coefficient versus mesh cells per chord.}
        \label{fig:ClNaca}
    \end{subfigure}
    \begin{subfigure}[b]{0.49\textwidth}
        \centering
        \includegraphics[width=\linewidth]{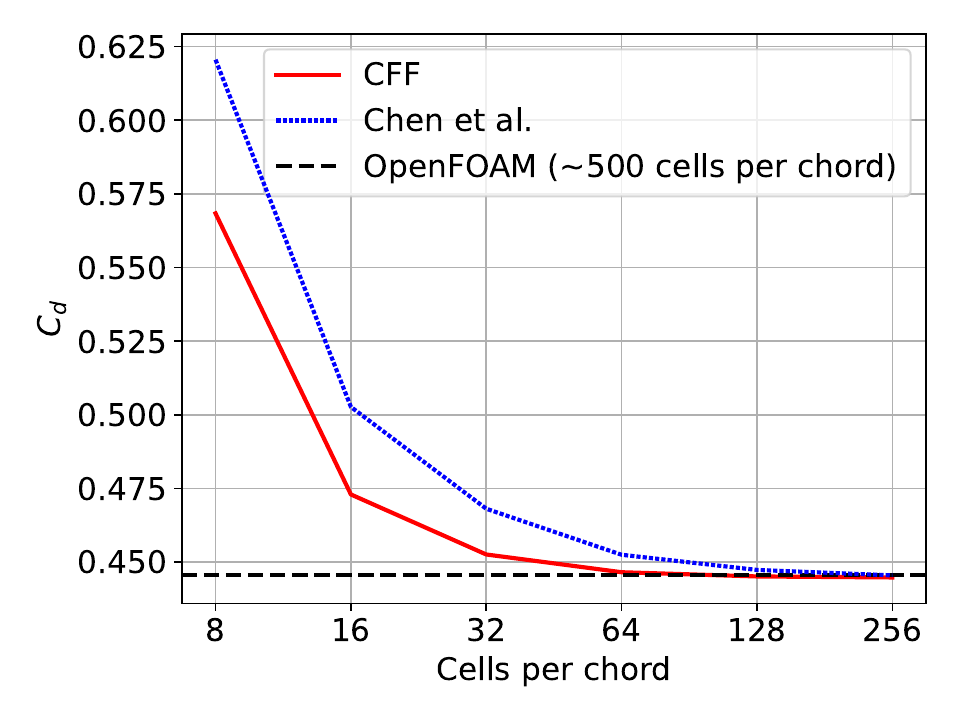}
        \caption{Drag coefficient versus mesh cells per chord.}
        \label{fig:CdNaca}
    \end{subfigure}
    \begin{subfigure}[b]{0.49\textwidth}
        \centering   
        \includegraphics[width=\linewidth]{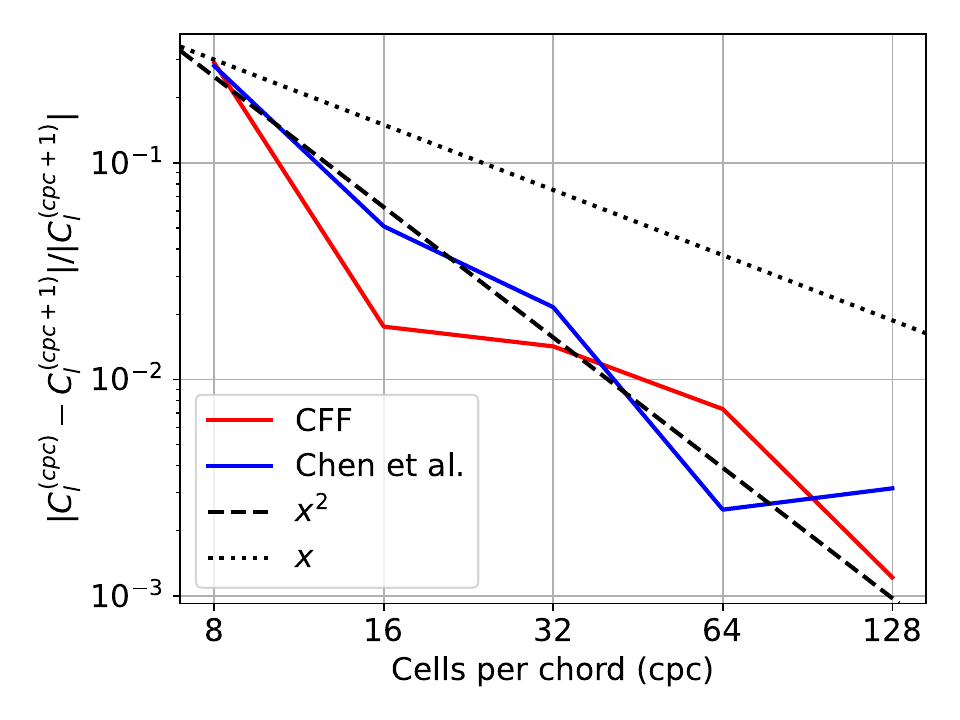}
        \caption{Lift coefficient relative error versus mesh cells per chord.}
        \label{fig:ClConvergenceNaca}
    \end{subfigure}
    \begin{subfigure}[b]{0.49\textwidth}
        \centering
        \includegraphics[width=\linewidth]{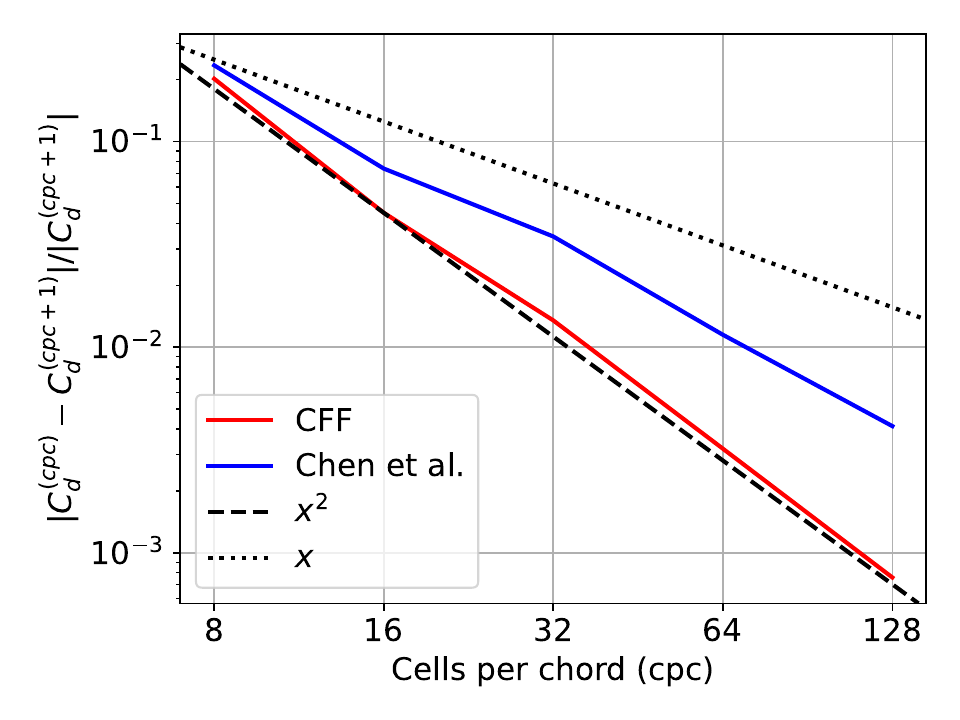}
        \caption{Drag coefficient relative error versus mesh cells per chord.}
        \label{fig:CdConvergenceNaca}
    \end{subfigure}
    \begin{subfigure}[b]{0.49\textwidth}
        \centering
        \includegraphics[width=\linewidth]{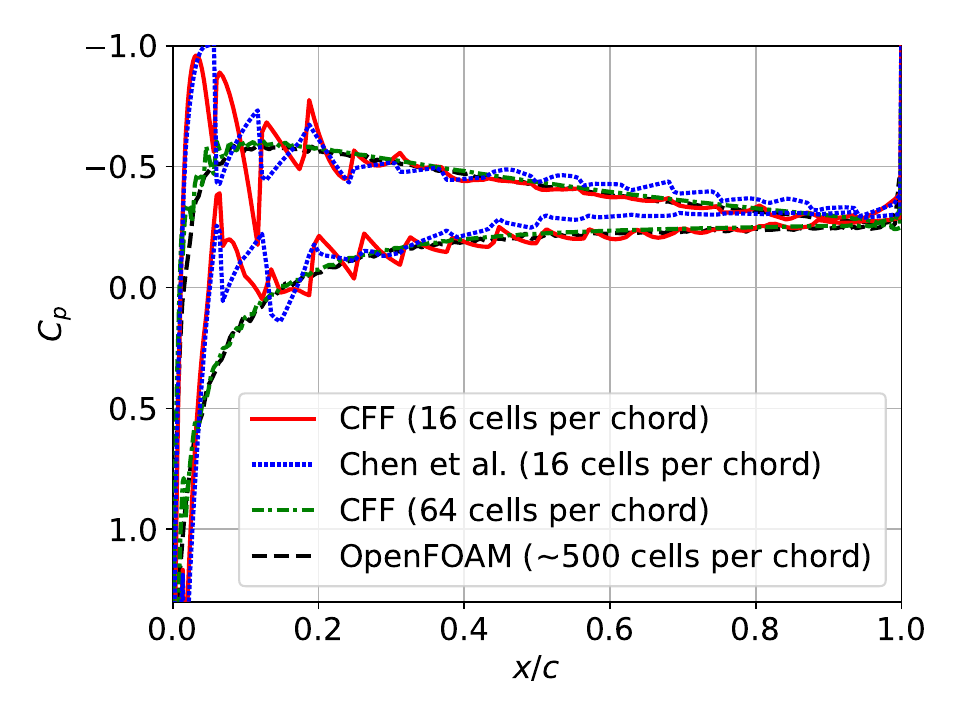}
        \caption{Pressure coefficient distribution.}
        \label{fig:CpNACA}
    \end{subfigure}
    \begin{subfigure}[b]{0.49\textwidth}
        \centering
        \includegraphics[width=\linewidth]{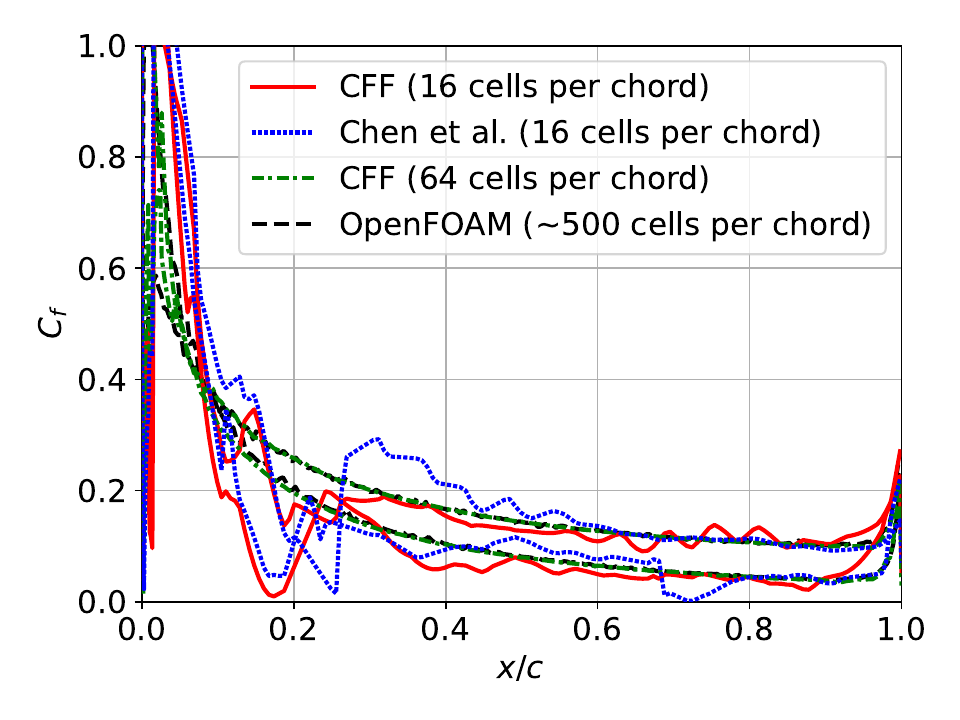}
        \caption{Skin friction coefficient distribution.}
        \label{fig:CfNACA}        
    \end{subfigure}
    \caption{Comparison of aerodynamic coefficients for flow over a NACA0012 airfoil at 5° angle of attack and Re = 100. Results from the proposed boundary treatment are compared against Chen et al. \cite{Chen1998b}, a fine OpenFOAM case is used as the ground truth.}
    \label{fig:airfoilResults}
\end{figure*}

We ran CFF as well as our implementation of Chen et al.'s algorithm for various grid refinements. The resulting lift and drag coefficients are plotted against the number of cells per chord in Fig.~\ref{fig:ClNaca} and Fig.~\ref{fig:CdNaca}, respectively. As the grid is refined, both algorithms converge to the OpenFOAM solution. For the lift coefficient, both algorithms perform similarly. However, for the drag coefficient, CFF performs significantly better. We note that with 16 cells per chord, CFF predicts both the lift and drag coefficient to within 5\% accuracy. To demonstrate the order of convergence, the relative error between successive grid refinements for the lift and drag coefficient is plotted in Fig.~\ref{fig:ClConvergenceNaca} and Fig.~\ref{fig:CdConvergenceNaca}, respectively. Chen converges with second order for the lift coefficient, but only with order $\sim$1.5 for the drag coefficient. CFF converges with second order for both.

The pressure and skin friction coefficient for the coarse case with 16 cells per chord are plotted along the chord in Fig.~\ref{fig:CpNACA} and Fig.~\ref{fig:CfNACA}, respectively. A slight discrepancy in $C_p$ is observed near the trailing edge compared to OpenFOAM. This is attributed to the difference in outflow condition. Compared to Chen, CFF produces better predictions of both surface force coefficients. 

For both algorithms, oscillations in the surface forces are observed. These are especially pronounced near the leading edge. Such oscillations are typical for methods based on Cartesian grids \cite{dePrenter2018}. Note that 16 cells per chord is much coarser than is typically used, so CFF results with 64 cells per chord are also presented. Indeed, with mesh refinement, the amplitude of these oscillations decreases significantly (in concert with the overall second-order rate of convergence). 

\begin{figure}
    \centering
    \includegraphics[width=\linewidth]{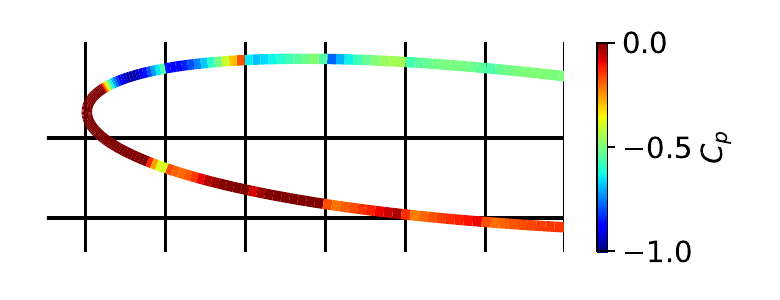}
    \caption{Front of the airfoil colored by pressure coefficient and shown with respect to the lattice cells, 16 cells per chord.}
    \label{fig:colorizedCp}
\end{figure}

\begin{figure}
    \centering
    \includegraphics[width=0.9\linewidth]{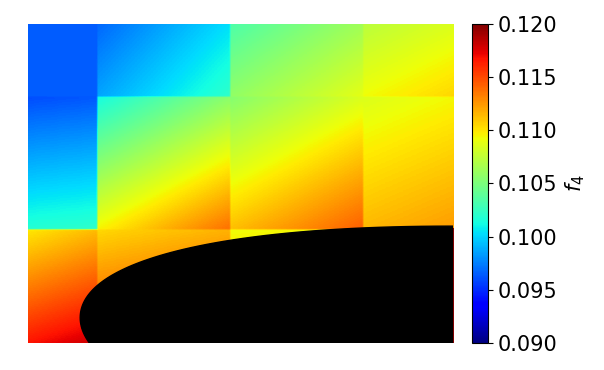}
    \caption{Discontinuous piecewise linears of downmoving populations near the NACA0012 leading edge, 16 cells per chord.}
    \label{fig:piecewisePolynomialsNaca}
\end{figure}

To further investigate these oscillations, the front of the airfoil colorized by $C_p$ is shown along with the lattice cells in Fig.~\ref{fig:colorizedCp}. Discontinuities in $C_p$ occur exactly at the cell borders. We hypothesize that the oscillations are the result of the discontinuous bases, employed by both CFF and Chen. This is supported by Fig.~\ref{fig:piecewisePolynomialsNaca}, where CFF's discontinuous piecewise linears of the downmoving populations ($f_4$) are shown near the leading edge. Discontinuities between the piecewise linears seem to originate from large variations between neighbouring cells resulting from too little resolution.

\section{Computational cost analysis}\label{sec:compCost}
The computational cost of a boundary treatment should be sufficiently small, to allow its use in practical calculations. In the current context, we distinguish two types of cost:
\begin{itemize}
    \item \textbf{Setup cost}: due to the processing of the geometry to obtain simple streaming rules. These are encoded in a matrix similar to the streaming matrix, as introduced in Sec.~\ref{subsec:conceptEigenValues}.
    \item \textbf{Cost per timestep}: due to the enforcement of the boundary conditions by applying the streaming rules at each timestep.
\end{itemize}
Note that if the geometry moves relative to the grid, the setup cost is incurred every timestep. We now analyze both types of cost for the present boundary treatment. Finally, we analyze the memory requirements.

\subsection{Setup cost}
The setup consists of three steps:
\begin{enumerate}
    \item Calculate the volume and the centroid of each cut cell. The associated cost depends on the number of cut cells as well as the number of facets:
    \begin{equation}
        C_{\text{step1}} = c_\alpha n_{\text{cut\_cells}} + c_\beta n_{\text{facets}},
        \label{eq:costStep1}
    \end{equation}
    where $c_\alpha$ and $c_\beta$ are constants.
    \item Calculate the coefficients of the discontinuous piecewise linears. This is done for each pgram-overlapped cell. Since information from step 1 can be reused, the cost at this step depends only on the number of pgram-overlapped cells. This in turn depends linearly on the number of cut cells, so the cost of this step scales as: 
    \begin{equation}
        C_{\text{step2}} = c_\alpha n_{\text{cut\_cells}},
        \label{eq:costStep2}
    \end{equation}
    where $c_\alpha$ is a constant, different from the $c_\alpha$ before.  
    \item Integrate the discontinuous piecewise linears over the subzones and collect the results in a sparse, global boundary matrix. The number of integrals that need to be calculated depends on the number of facets. Note that facets are further cut down within each cell before computing the integrals (see Sec.~\ref{subsec:MappingAndIntegration}). This means that the cost of this step scales as:
    \begin{equation}
        C_{\text{step3}} = c_\alpha n_{\text{cut\_cells}} + c_\beta n_{\text{facets}},
        \label{eq:costStep3}
    \end{equation}
    where $c_\alpha$ and $c_\beta$ are constants, different from those before.
\end{enumerate}
Combining Eq.~\ref{eq:costStep1}--\ref{eq:costStep3}, we arrive at a total setup cost of:
\begin{equation}
    C_{\text{setup}} = c_\alpha n_{\text{cut\_cells}} + c_\beta n_{\text{facets}},
\end{equation}
where $c_\alpha$ and $c_\beta$ are constants, representing the sum of those before. Note that this cost is only incurred once for stationary geometries.

\subsection{Cost per timestep}
In the setup, simple streaming rules were derived based on the geometry and stored in a sparse, global boundary matrix. For each timestep, the boundary condition is enforced by multiplying the boundary matrix with the global vector of populations, resulting in updated populations. This means that the timestep cost of one population in one cell depends on the number of nonzeros in the corresponding boundary matrix row.


We now derive the theoretical maximum number of nonzeros in a row of the boundary matrix. Each nonzero corresponds to either a direct stream or a reflection. Directly streamed populations can originate from only one cell. Reflected populations, on the other hand, can originate from up to seven cells (as is the case for a diagonal stream). This means that for the schemes of Chen et al. \cite{Chen1998b} and Li et al. \cite{Li2004}, the maximum number of nonzeros per row is eight. For CFF, this maximum is larger, because populations are represented using  discontinuous piecewise linears, which in turn are constructed from the eight neighbouring cells. Taking this into account gives a theoretical maximum of 33 nonzeros per row for CFF.

To verify these theoretical estimates, we report statistics of the observed number of nonzeros per row in the boundary matrix of the NACA0012 geometry, described in Sec.~\ref{subsec:airfoilResults}. The average and maximum values are listed in Tab.~\ref{tab:boundaryNNZ}. These results confirm our theoretical estimates. Moreover, the average number of nonzeros gives insights into the actual cost. The timestep cost of CFF is approximately two times larger than the boundary treatment of Chen et al. Furthermore, $\sim$10 nonzeros means that an average of 19 operations (10 multiplications, 9 additions) are needed per timestep to enforce the boundary condition for one population. 


\begin{table}
    \centering
    \caption{Statistics for the number of nonzeros on a row of the boundary matrix.}
    \label{tab:boundaryNNZ}
    \begin{tabular}{|c|c|c|c|} \hline
    & \makecell{Theoretical\\ maximum} & \makecell{NACA0012\\ maximum\\ (measured)} & \makecell{NACA0012\\ average\\ (measured)} \\ \hline
    CFF & 33 & 31 & 10.0 \\ \hline
    Chen et al. \cite{Chen1998b} & 8 & 7 & 4.4 \\ \hline
    \end{tabular}
\end{table}

Now that we have estimated the timestep cost per pgram-overlapped cell, the total timestep cost is estimated. Using that the number of pgram-overlapped cells scales proportionally with the number of cut cells, the timestep cost scales as:
\begin{equation}
    C_{\text{timestep}} = c_\alpha n_{\text{cut\_cells}},
    \label{eq:timestepCost}
\end{equation}
where $c_\alpha$ is a constant, different from the $c_\alpha$ before. In our experience, this cost is usually lower than that of the standard streaming and collision in the bulk. Specifically when refining grids, as the number of bulk voxels grows with a higher rate than the number of cut voxels.

Finally, Eq.~\ref{eq:timestepCost} implies that the timestep cost does not depend on the number of facets. Thus, for stationary geometries, a fine surface discretization can be used to minimize the associated error, while keeping the same cost per timestep. However, the setup cost would increase. 

\subsection{Memory requirements}
The sparse boundary matrix is expected to dominate memory usage related to boundary handling. It has a certain number of nonzeros per row (independent of grid refinement), and the number of rows scales linearly with the number of cut cells. As a result, the memory required by the boundary matrix scales as: 
\begin{equation}
    M_{\text{boundary\_matrix}} = c_\alpha n_{\text{cut\_cells}}
\end{equation}
where $c_\alpha$ is an arbitrary constant, different from the $c_\alpha$ before. In our experience, the required memory is typically smaller than that of the bulk.

\section{Conclusion and recommendations}
In this work, a higher-order volumetric-type boundary treatment for the lattice-Boltzmann method was introduced and tested in 2D. Also, we introduced a novel analysis tool for verifying the stability of boundary treatments, based on eigenvalues of the streaming step including enforcement of boundary conditions.

The accuracy of the boundary treatment was measured via mesh convergence studies on a quasi-1D channel and a 2D channel inclined with respect to the lattice. Convergence rates of $\geq2$ were observed for velocity, which is an improvement over previous work \cite{Li2012}. In order to isolate the effect of the boundary treatment, we introduced a case with pure advection (only streaming), allowing us to study the convergence rate of the populations themselves. In both 1D and 2D, we found a convergence rate of approximately 3 for each population.

Finally, we tested the implementation on a NACA0012 airfoil at $5^\circ$ angle of attack and Reynolds number 100. The lift and drag coefficients were predicted more accurately than precursor volumetric approaches at the same resolution.

We have presented a promising boundary treatment algorithm. To apply it to industrial cases, some extensions need to be considered. Specifically: Nonconvex geometries, for which particles may advect from one facet to another within one timestep \cite{Chen1998b}; 3D geometries, where cutting the facets and obtaining integrals becomes much more involved; non-uniform meshes, which require special treatment when a pgram overlaps two refinement levels.

\section{Acknowledgments}
The authors gratefully acknowledge financial support from the Netherlands Organisation for Scientific Research (NWO) through grant VENI-20153.

\section{Data availability}
The datasets and source code supporting the findings of this study are openly available on GitLab at: https://gitlab.tudelft.nl/knhoefnagel/convexcff2d


\nocite{*}

\bibliography{bibliography}


\end{document}